\documentclass[manuscript,screen]{acmart} 

\AtBeginDocument{%
  }

\setcopyright{none}
\copyrightyear{2026}
\acmYear{2026}
\acmDOI{10.1145/3811815} 

\usepackage{environment/031421AkiEnSty}

\begin{document}

\title{Optimal design of lottery with cumulative prospect theory}

\author{Shunta Akiyama}
\authornote{Both authors contributed equally to this research.}
\email{miagtoh1997@gmail.com}
\orcid{0009-0009-3550-7921}
\author{Mitsuaki Obara}
\authornotemark[1]
\email{polyelo2plumo@gmail.com}
\orcid{0000-0002-3731-6776}
\author{Yasushi Kawase}
\email{yambian@gmail.com}
\orcid{0000-0001-5626-779X}
\affiliation{%
  \institution{The University of Tokyo}
  \city{Hongo}
  \state{Tokyo}
  \country{Japan}
}


\begin{abstract}

Lotteries are a prevalent form of gambling between a seller and buyers. Designing a lottery requires a model of how buyers make decisions when confronted with uncertain outcomes. 
\textit{Cumulative prospect theory (CPT)} is a descriptive model that captures people's propensity to overestimate extreme events and their different attitudes toward gains and losses. 
In this study, we design a lottery that maximizes the seller's profit when the buyers' decision-making adheres to the CPT framework. The main difficulty is the nonconvexity of the CPT framework, which we overcome by reformulating the problem as a three-level optimization problem and characterizing its optimal solution. Based on the analysis, we propose a linear-time algorithm that computes the optimal lottery. Furthermore, we present an efficient algorithm applicable to a broader setting with a ticket price constraint. 
This is the first study to employ the CPT framework in designing an optimal lottery with more than two outcomes.


\end{abstract}



\received{29 January 2025}
\received[revised]{28 January 2026}
\received[accepted]{14 April 2026}

\maketitle

\section{Introduction}
\noindent
Lotteries are a prevalent form of gambling involving a seller and multiple buyers. 
In modern lotteries, the seller, typically a government agency, announces the price per ticket and the prizes in advance.
Buyers then purchase tickets if they perceive the value of the ticket to be worth the expense.
Prizes of random amounts are awarded to buyers according to predetermined rules.

The seller profits from the difference between the sales revenue and the total prizes awarded. 
Thus, devising a profitable lottery design is of primary importance to the seller. 
From this perspective, the optimal design of lotteries has been explored in previous studies~\citep{maeda2008optimal,DenneryDirer2014optimlottery}.
To achieve the optimal design, it is necessary to model the conditions under which a buyer purchases a lottery ticket with uncertain outcomes.

Several models have been proposed to explain decisions under uncertainty.
The expected utility hypothesis (EU) is a seminal framework~\citep{neumann1944theory,friedman1948utility,fishburn1970utility}
that introduces a utility function to describe people's satisfaction with outcomes. 
Agents choose actions based on the subjective utility, that is, the sum of the products of the utilities of outcomes and their corresponding probabilities.
However, EU does not satisfactorily explain the behavior of people buying lottery tickets~\citep{Quiggin1991optimlotterydesign}. 

This study employs \textit{cumulative prospect theory (CPT)}, an influential model derived from EU by \citet{tversky1992advances}.
CPT incorporates a cumulative manner to calculate decision weights and introduces sign-dependent preferences. 
These enhancements give CPT desirable properties, such as first-order stochastic dominance ~\citep{FennemaWakker1997OrgCPTDiscussion}.

\subsection{Our contributions} 
This paper focuses on designing a profit-maximizing lottery for the seller, assuming buyers follow CPT.
Developing characterizations and computational methods for optimal mechanisms is fundamental in operations research, computer science, and economics.

We first formulate the problem of finding an optimal lottery that maximizes the seller's profit when the buyers follow CPT.
To the best of the authors' knowledge, this is the first study that employs the CPT framework to design an optimal lottery.
The derived problem is nonconvex and constrained; hence, it is challenging to characterize or compute its optimal solution. 
To overcome this difficulty, we reformulate the problem as a three-level optimization problem. 
By using this reformulation, we reveal that the prizes of the optimal lottery are divided into three parts: a polynomially increasing part, a fixed amount higher than the ticket price, and no prizes.
Based on the result, we first present an algorithm that computes the characterization parameters in quadratic time with respect to the number of lottery tickets. Then, we improve the computational time to linear by further investigating the structure of the optimal lottery.
Such an efficient algorithm is necessary when we design large-scale lotteries that appear in practice, such as \textit{Powerball}.

We also analyze the lottery optimization problem with an additional constraint on the ticket price, which is more useful in practice. 
In this case, it is difficult to obtain the optimal lottery analytically. 
However, we characterize it as the optimal solution of a 1-dimensional optimization problem, which can be solved efficiently by numerical algorithms such as the Newton method. 
Then, we provide an algorithm that outputs the optimal lottery by solving a 1-dimensional optimization problem quadratic times.
Moreover, we give a linear-time algorithm that finds the optimal lottery in a specific case.

Finally, we conduct numerical experiments to illustrate the optimal structures of lotteries and confirm that the algorithms above compute optimal lotteries within a practical time.

\subsection{Related studies}
Several other frameworks stem from EU. 
Rank-dependent expected utility~(RDEU) is an improvement of EU~\citep{Quiggin1982RankDependentEU}, successfully capturing the overestimation of an unlikely extreme event, such as a jackpot in a lottery.
Mathematically, the utility in this model is calculated by using decision weights, which are probabilities distorted by an inverse S-shaped probability weighting function in a cumulative manner.
Another series of studies is prospect theory (PT) in behavioral economics~\citep{KahnemanTversky1979prospecttheory,starmer2000developments,barberis2013thirty}. 
PT considers sign-dependent preferences, where risk attitudes depend on whether the individual evaluates a gain or a loss.
Here, the utility is defined by an S-shaped value function instead of the utility functions above, together with distorted probability. 
CPT improves RDEU and PT by incorporating cumulative decision weight calculations from RDEU and sign-dependent preferences from PT.


Previous studies have investigated the optimal lottery design mostly based on RDEU or other frameworks. 
There are several other differences between this study and previous studies: \citet{Quiggin1991optimlotterydesign} dealt with maximizing the buyer's utility under a fixed profit for the seller with RDEU. 
The profit maximization was studied in \cite{maeda2008optimal,DenneryDirer2014optimlottery}.
\citet{maeda2008optimal} analyzed an optimal design based on the utility framework where the number of $n$-th winning lottery is determined to be $n$. While he provided a closed-form solution for a specific type of lottery, our study does not require such a pre-specification and optimizes the entire structure of the lottery based on CPT.
\citet{DenneryDirer2014optimlottery} optimized the probability of each prize in addition to its value and the number of prizes. 
They concluded that their model is incompatible with a discrete number of prizes.
The comparison between our setting and existing ones is summarized in \cref{table:comparisonexistingworks}.

As for the lottery design based on CPT, \citet{azevedo2012risk} analyzed the seller's profit against CPT agents with two lottery tickets, namely, one winning ticket and one losing ticket.
They derived that the seller can extract arbitrarily large profits from the agent, raising a pessimistic result for lottery design in the CPT framework.
However, this result is attributable to the restricted structure of CPT in their model, where the value functions for positive and negative outcomes differ only by a constant factor.
Our study investigates the optimal lottery problem with multiple tickets based on the full CPT framework, allowing relationships between the value functions beyond just a constant factor.
As a result, we successfully derive meaningful cases where the optimal lottery consists of finite lottery awards and a specific ticket price.
Note that, since our setting covers that of \cite{azevedo2012risk}, our optimal lottery likewise encompasses cases where the seller can extract arbitrarily large profits.
We also clarify the conditions under which such situations occur.

Analysis with CPT becomes more challenging than that with RDEU.
For example, consider the optimal design of insurance, which partly shares similarities with that of lotteries in that both problems involve fixed payments and random rewards.
Although several optimal designs of insurance have been investigated using RDEU~\citep{Bernardetal2015optiminsuranceunderRDEU,xu2019optimal,ghossoub2019optimal}, the design using CPT is still limited to a simplified CPT setting~\citep{Sungetal2011behavioraloptiminsurance}.
One of the reasons is the (inverse) S-shape of the probability weighting function and the value function. 
As \citet{Bernardetal2015optiminsuranceunderRDEU} stated, the inverse S-shape often causes analytical difficulty due to the nonconvexity.
The lottery problem in this study also encounters the same difficulty: the S-shape results in nonconvexity.
Nevertheless, we resolve it by decomposing the lottery design problem into three-level optimization.
Although the variables to be optimized in the design of insurance differ from those in the design of lotteries, we believe that our technique is also helpful in designing optimal insurance.
    
\renewcommand{\checkmark}{\textcolor{green!80!black}{\ding{51}}}
\newcommand{\bcheckmark}{\textcolor{green!80!black}{\ding{52}}}
\newcommand{\xmark}{\textcolor{red}{\ding{55}}}
\begin{table}[t]
\caption{Comparing results on the lottery design}
\centering
\begin{tabular}{l|ccc}
\hline
                                     & Framework        & Characterization
                                     & Computation 
                                     \\ \hline
\citet{Quiggin1991optimlotterydesign}& RDEU         & \checkmark          & \xmark       \\
\citet{DenneryDirer2014optimlottery} & RDEU         & \checkmark          & \xmark       \\
\citet{maeda2008optimal}             & EU-based 
& \checkmark          & \checkmark   \\
\textbf{This work}                   & \textbf{CPT} & \bcheckmark         & \bcheckmark  \\ \hline
\end{tabular}\label{table:comparisonexistingworks}
\end{table}

Another line of recent research~\citep{chawla2018revenue,liu2022risk,chawla2020menu,chawla2019buy} explores the mechanism design with prospect theoretic buyers (including CPT), which can be interpreted as a variant of the lottery optimization problem.
These papers investigate allocation schemes that achieve the maximum profit when items with certain values are given.  
Among them, \citet{chawla2018revenue} studied revenue maximization with prospect-theoretic buyers and showed that the revenue-optimal mechanism can be represented as a menu of two-outcome lotteries, highlighting structural properties of revenue-optimal mechanisms.
Their work does not fully utilize the CPT structure, i.e., the S-shaped probability weighting function and the value function. 
In contrast, our result characterizes both the seller's profit-maximizing lottery outcomes and their values under the full CPT framework.

\section{Preliminaries}\label{sec:preriminaries}
    
    Let $\setR[]$ be the set of real numbers.
    We will denote by $\setRp$ (resp., $\setRn$) the set of nonnegative (resp., nonpositive) real numbers.
    Let $\setN\coloneqq\brc*{1,2,\ldots}$ be the set of natural numbers.
    For $n\in\setN$, we denote the set $\brc*{1,\ldots,n}$ by $\setIntvl[n]$.
    We define $\setNz\coloneqq \setN\cup\brc*{0}$.
    For integers $a$ and b, if $a > b$, the set $\brc*{i\in\mathbb{Z}\mid a\le i\le b}$ is understood to be empty.
        
\subsection{Cumulative prospect theory}
    Let us consider prospects $\paren*{\wconb[1],\p[1]},\ldots,\paren*{\wconb[\N],\p[\N]}$, 
    where $\wconb[\idx]\in\setR[]$ is a potential outcome and $\p[\idx]~\paren*{\ge 0}$ is the probability of yielding it. 
    We assume that the outcomes are arranged in the nondecreasing order (i.e., $\wconb[1] \leq \wconb[2]\leq\dots\leq\wconb[\N]$) and $\sum_{\idx=1}^{\N}\p[\idx]=1$ holds.
    We call a nonnegative outcome a \textit{gain} and nonpositive one a \textit{loss}.
    We interpret zero as either a gain or a loss.
    
    In the CPT framework~\citep{tversky1992advances}, an agent makes a decision according to the CPT utility of the prospects.
    To define the CPT utility, we first introduce two critical ingredients, that is, a \textit{value function} and \textit{decision weights}.

\subsubsection{\textbf{Value function}}
    A value function represents an agent's preference for outcomes.
    It is concave and convex for gains and losses, respectively; see \cref{fig:utlity}.
    The concavity represents the risk aversion of an agent when facing risk related to gains, and the convexity represents risk-seeking behavior when facing losses.
    
    The most widely used value function for CPT is
    \begin{align}
    \label{def:utilityfunc}
        \U\paren*{\wconb} \coloneqq
        \begin{cases}
            \wconb^{\alpha} & \paren*{\wconb \geq 0}, \\
            -\propminus \paren*{-\wconb}^{\beta} & \paren*{\wconb < 0},
        \end{cases}
    \end{align}
    where $\wconb \in \setR[]$ is an outcome and $\alpha,\beta,\lambda$ are hyperparameters satisfying $\alpha,\beta\in\paren*{0,1}$ and $\lambda>0$.
    The parameters $\paren*{\alpha, \beta, \propminus}$ are usually set for the function to be steeper for losses than gains, which is known as loss aversion.
    Throughout this paper, we assume that the value function is represented as \cref{def:utilityfunc}.
    We will discuss possible extensions to more general value functions in \cref{sec:conclusion}.

\subsubsection{\textbf{Decision weights}}
\begin{figure}[htbp]
      \begin{minipage}[t]{0.49\linewidth}
        \includegraphics[width=\linewidth]{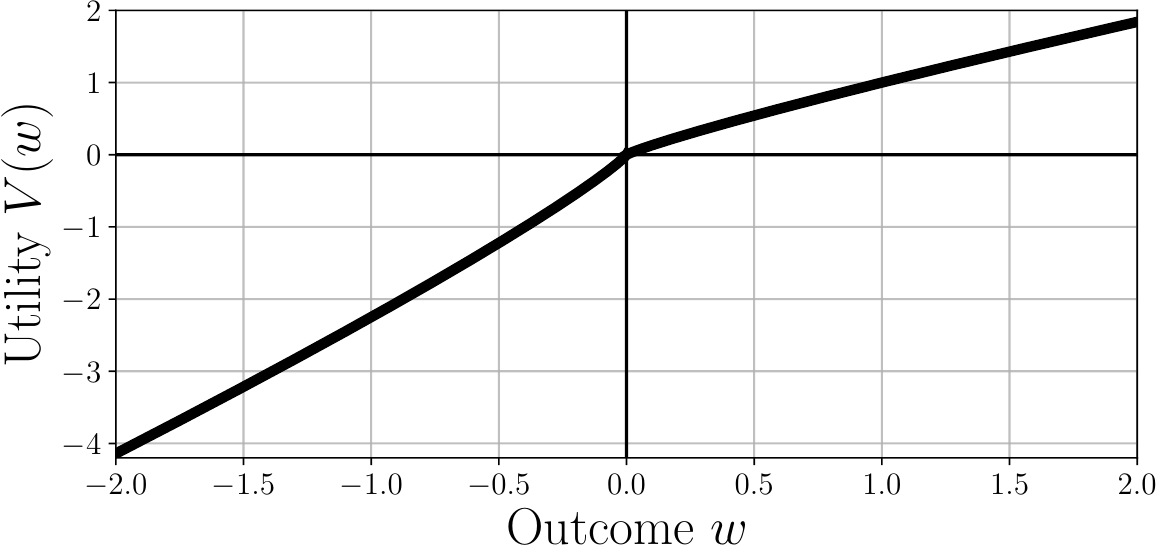}
        \caption{Value function ($\alpha=0.88$, $\beta = 0.88$, $\propminus=2.25$)}
        \label{fig:utlity}
      \end{minipage}\hfill
      \begin{minipage}[t]{0.49\linewidth}
        \includegraphics[width=\linewidth]{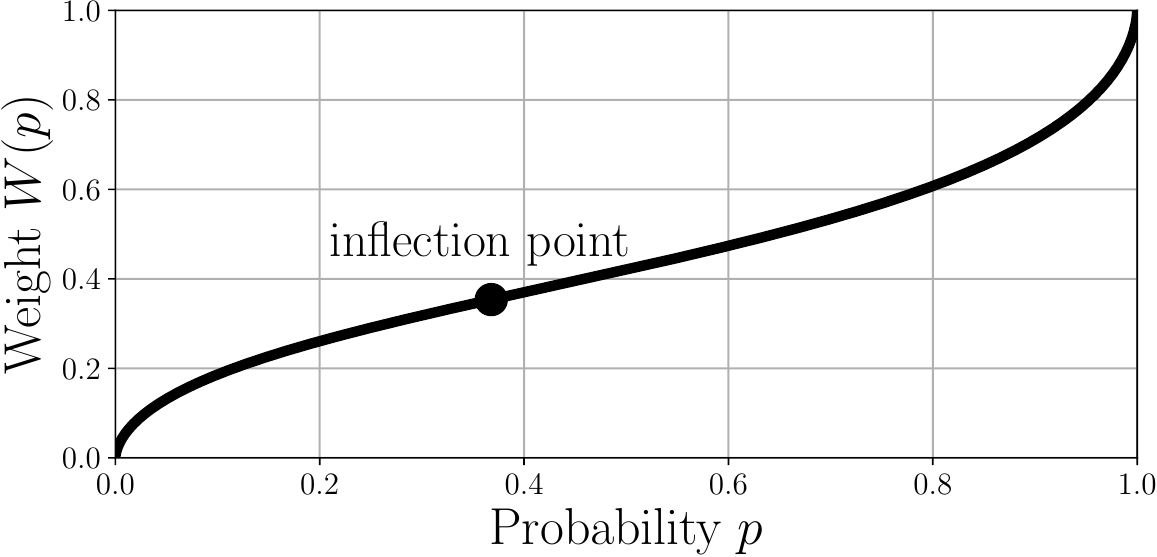}  
        \caption{Probability weighting function ($\weightfuncparamplus = 0.61$)}
        \label{fig:pwf}
      \end{minipage}
    \end{figure}

    Decision weights represent an agent's subjective probabilities of the corresponding outcomes. 
    We say that $f\colon \sbra*{0,1}\to \setR$ is an \textit{inverse S-shaped} function if it is strictly increasing, continuously differentiable and there exists $x_0\in \sbra*{0,1}$ such that $x\mapsto f^{\prime}\paren*{x}$ is strictly decreasing on $\sbra*{0, x_0}$ and strictly increasing on $\sbra*{x_0,1}$. 
    We call such $x_{0}$ the \textit{inflection point}; see \cref{fig:pwf}.
    Note that $x_0$ satisfies $f^{\prime\prime}\paren*{x_0} = 0$ if $f$ is twice differentiable.
    Let $\weightfuncplus,\, \weightfuncminus\colon\sbra*{0,1}\to\sbra*{0,1}$ be inverse S-shaped functions satisfying $\weightfuncplus\paren*{0} = \weightfuncminus\paren*{0} = 0$ and $\weightfuncplus\paren*{1} = \weightfuncminus\paren*{1} = 1$.
    We call such $\weightfuncplus$ and $\weightfuncminus$ \textit{probability weighting functions} for gains and losses, respectively. 
    Typical probability weighting functions for gains and losses are 
    \begin{align}
       \weightfunc\paren*{\p} {\coloneqq} \tfrac{\p^{\weightfuncparam}}{\paren*{\p^{\weightfuncparam} + \paren*{1-\p}^{\weightfuncparam}}^{\frac{1}{\weightfuncparam}}}
       \quad \text{and}\quad
       \weightfuncminus\paren*{\p} {\coloneqq} \tfrac{\p^{\weightfuncparamminus}}{\paren*{\p^{\weightfuncparamminus} + \paren*{1-\p}^{\weightfuncparamminus}}^{\frac{1}{\weightfuncparamminus}}}
       \label{eq:weightfunc}
    \end{align}
    with $\weightfuncparam, \weightfuncparamminus \in (0,1]$, which are given by \citet{tversky1992advances}.
    
    Let $\numplus$ and $\numminus$ be the numbers of nonnegative and nonpositive outcomes, respectively; that is, it follows that $\wconb[\numminus]\leq 0$, $\wconb[\numminus+1]\geq 0$, and $\numminus+\numplus=\N$. 
    Then, the decision weights are defined as
    \begin{align}
        &\hminus[\idxminus] \coloneqq \weightfuncminus \paren*{\sum_{\idx=1}^{\idxminus} \p[\idx]} - \weightfuncminus\paren*{\sum_{\idx=1}^{\idxminus-1} \p[\idx]} \quad \paren*{\idxminus \in \setIntvl[\numminus]},\label{eq:hminus}\\
        &\hplus[\idxplus] \coloneqq \weightfuncplus\paren*{\sum_{\idx=\idxplus}^{\numplus} \p[\idx]} - \weightfuncplus\paren*{\sum_{\idx=\idxplus+1}^{\numplus} \p[\idx]} \quad \paren*{\idxplus \in \setIntvl[\numplus]}.\label{eq:hplus}
    \end{align}
    Here, we regard the empty sum as zero.
    Note that the inverse S-shape of $\weightfuncplus$ and $\weightfuncminus$ implies the overestimation of outcomes with large losses or gains; that is, the decision weights of such extreme events tend to be larger than those of intermediate ones. 
    
    
    
\subsubsection{\textbf{CPT utility}}
    Let $\wminus \coloneqq \trsp{\paren*{\wconb[1], \ldots, \wconb[\numminus]}} \in \setRn[\numminus]$ be the first $\numminus$ components of $\w$ that correspond to the losses for the agent.
    We also define the gains as $\wplus \coloneqq \trsp{\paren*{\wconb[\numminus + 1], \ldots, \wconb[\N]}} \in \setRp[\numplus]$.
    Then the CPT utility of the agent is defined as
    $\sum_{\idxminus=1}^{\numminus}\hminus\cdot\U\paren*{\wminus[\idxminus]} + \sum_{\idxplus=1}^{\numplus}\hplus\cdot\U\paren*{\wplus[\idxplus]}.$
    Agents make decisions according to their expected utilities based on their subjective value and probability. 
    We emphasize that the CPT utility is different from the expected value.
    \begin{example}
        Let us consider the lottery with the outcome $-0.1$ with $90\%$, $0.1$ with $9.5\%$, and $10$ with $0.5\%$.
        We can confirm that the expected value of this lottery is $-0.0305$ ($=-0.1\cdot 0.9+0.1\cdot 0.095+10\cdot 0.005$).
        On the other hand, the CPT utility with the parameters $\paren*{\alpha,\beta,\propminus,\weightfuncparamplus, \weightfuncparamminus}=\paren*{0.42, 0.83, 1.62, 0.44, 0.60}$ is about $0.0953$ 
        ($=V\paren*{-0.1}\cdot\weightfuncminus\paren*{0.9}+V\paren*{0.1}\cdot\paren*{\weightfuncplus\paren*{0.095+0.005}-\weightfuncplus\paren*{0.005}}+V\paren*{10}\cdot\weightfuncplus\paren*{0.005}$).
        This implies that even if the expected value is negative, the lottery can attract the agent with those CPT parameters.
    \end{example}
    
\section{Formulation}\label{sec:formulation}
    We formulate the optimal lottery design as an optimization problem in which the objective is to maximize the seller's profit, subject to the constraint that the lottery is attractive to buyers from the perspective of CPT.
    To ensure that the tickets are sold, we require that the CPT utility of a lottery ticket must be nonnegative for buyers.
    Importantly, this requirement does not limit buyers from purchasing multiple tickets.
    We will also explore the scenario in which the CPT utility is at least a certain positive value in \cref{sec:conclusion}. 
    
    Note that the nonnegative condition can be interpreted as \textit{individual rationality} in the auction theory, and our problem can be viewed as a \textit{revenue maximizing auction}~\citep{myerson1981optimal,roughgarden2016twenty,chawla2018revenue,liu2022risk} for no items. Our findings represent a novel contribution even within this auction context as well.

    We consider a scenario where all the tickets are sold, assuming a sufficient number of potential buyers. 
    Additionally, we assume homogeneity among the buyers; that is, we consider the optimal lottery with a value function $\U$ and a pair of probability weighting functions $\weightfuncplus$ and $\weightfuncminus$, representing the CPT utility in the target population.
    This assumption of agent homogeneity is commonly made in previous studies that model agents' behavior using the CPT framework~\citep{tversky1992advances,rieger2011prospect,Riegeretal2017estimateCPTparams}. 
    While this assumption may not perfectly capture the intricacies of reality, where decision-making can vary even within the same personas, it remains a valuable tool for gaining clear insights and facilitating a more tractable analysis of the problem.


    Let $\N \in \setN$ be the number of the lottery tickets. 
    In addition, let $u \coloneqq \trsp{\paren*{\wconb[1], \ldots, \wconb[\N]}} \in \setR[\N]$ be the outcomes of tickets, i.e., the differences between lottery prizes and the ticket price.
    Specifically, the ticket price is a real number $q$ satisfying $q \ge -\min_{i=1}^{\N} \wconb[i]$, and the corresponding prize values are given by $\wconb[i] + q$ for $i = 1, \dots, \N$.
    We note that, under this formulation, the worst outcome can be naturally interpreted as an implicit ticket price, in the sense that $-\min_{i=1}^{\N} \wconb[i]$ represents the maximum payment a buyer may incur.
    Following the original CPT paper~\cite{tversky1992advances} and much of the subsequent literature, we assume that buyers evaluate their utility based on outcomes. 
    That is, rather than treating the ticket price and the prize amount separately, buyers perceive the outcome as $u$, the difference between the prize amount and the ticket price.
    
    We assume that the correspondence between the tickets and the outcomes is chosen uniformly at random from the $N!$ possibilities.
    Hence, the prospects of a ticket are $\paren*{\wconb[1],1/\N},\ldots,\paren*{\wconb[\N],1/\N}$.
    We also investigate non-uniform settings in \cref{sec:conclusion}. 
    As in \cref{sec:preriminaries}, we denote the numbers of nonnegative and nonpositive outcomes by $\numplus$ and $\numminus$, respectively.
    We denote the losses for the agent by $\wminus \coloneqq \trsp{\paren*{\wconb[1], \ldots, \wconb[\numminus]}} \in \setRn[\numminus]$, and the gains by $\wplus \coloneqq \trsp{\paren*{\wconb[\numminus + 1], \ldots, \wconb[\N]}} \in \setRp[\numplus]$. 
    We now consider the optimal design of a lottery that maximizes the seller's profit as follows:
    \begin{align}
        \label{prob:orgNminusplus}
            \max_{\numminus,\,\numplus\in \setN\cup\brc*{0}}~\optval\paren*{\numminus, \numplus}
        \quad \subjectto~\numminus + \numplus = \N.
    \end{align}
    Here, $\optval\paren*{\numminus, \numplus}$ is defined by the optimal value of the following problem 
    with fixed $\numminus$ and $\numplus$: 
    \begin{subequations}\label{prob:orgmaxprofit}
        \begin{align}
            \max_{\substack{\wminus\in\setR[\numminus],\wplus[]\in\setR[\numplus]}} \quad&{\sum_{\idxminus=1}^{\numminus} \paren*{-\wminus[\idxminus]} - \sum_{\idxplus=1}^{\numplus} \wplus[\idxplus] \protect\label{objfun:CPTmaxprofit}}\\
            \subjectto \quad&\sum_{\idxminus=1}^{\numminus}\hminus\cdot \U\paren*{\wminus[\idxminus]} +  \sum_{\idxplus=1}^{\numplus}\hplus\cdot \U\paren*{\wplus[\idxplus]}\ge 0, \label{cstr:CPTexpectedutility}\\
            \quad&0\leq \wplus[1] \leq \wplus[2] \leq \cdots \leq \wplus[\numplus]\protect\label{cstr:gainascending},\\
            \quad&\wminus[1]\leq \wminus[2] \leq \cdots \leq \wminus[\numminus]\le 0 \protect\label{cstr:lossascending},
        \end{align}
    \end{subequations}
    where $\hminus[\idxminus]~\paren*{\idxminus \in \setIntvl[\numminus]}$ and $\hplus[\idxplus]~\paren*{\idxplus \in \setIntvl[\numplus]}$ are defined as \eqref{eq:hminus} and \eqref{eq:hplus}, respectively.
    It is not difficult to see that $\optval\paren*{0,N}=\optval\paren*{N,0}=0$.
    Hence, in what follows, we only consider the case where $\numplus$ and $\numminus$ are positive.
    
    The objective function~\cref{objfun:CPTmaxprofit} is the seller's profit, the sum of the seller's sales and payment. 
    \cref{cstr:CPTexpectedutility} ensures that each buyer can always attain at least an equivalent amount of CPT utility from purchasing a ticket as from not purchasing one.
    Constraints~\cref{cstr:lossascending,cstr:gainascending} impose the signs and the orders on the outcomes. 
    We remark that the constraint \cref{cstr:CPTexpectedutility} is indeed nonconvex.

    By letting $\yminus=-\U\paren*{\wminus[\idxminus]}$ for $\idxminus \in \setIntvl[\numminus]$ and $\yplus=\U\paren*{\wplus[\idxplus]}$ for $\idxplus \in \setIntvl[\numplus]$, we provide a reformulation of the problem
    ~\cref{prob:orgmaxprofit} in the following proposition.
    \begin{proposition}
        \label{prop:equivalentproblem}
        The problem 
        \cref{prob:orgmaxprofit}
        is equivalent to the following optimization problem:
        \begin{align}
            \min_{\substack{\eqval\in\setRp[]}}\quad{\fminus_{\numminus}\paren*{v}+\fplus_{\numplus}\paren*{v},}\label{prob:middlevsubprob}
        \end{align}
    where $\fminus_{\numminus}\paren*{\eqval}$ is the optimal value of
    \begin{subequations}\label{cstr:4thlosssubprob}
        \begin{align}
            \min_{\substack{\yminus[]\in\setR[\numminus]}}\quad&{\sum_{\idxminus=1}^{\numminus} \U^{-1}\paren*{-\yminus[\idxminus]}\quad\paren*{\text{\scriptsize$=-\sum_{\idxminus=1}^{\numminus}\paren*{\frac{\yminus}{\propminus}}^{\frac{1}{\beta}}$}}}\label{objfun:4thloss}\\
            \subjectto \quad&\yminus[1]\geq \yminus[2] \geq \cdots \geq \yminus[\numminus]\geq 0,\protect\label{cstr:yminusMonotone}\\
            \quad&{\sum_{\idxminus=1}^{\numminus}\hminus\yminus=v}\label{cstr:sumyminus}
        \end{align}
    \end{subequations}
    and $\fplus_{\numplus}(v)$ is the optimal value of
    \begin{align}
        \begin{split}\label{cstr:4thgainsubprob}
            \min_{\substack{\yplus[]\in\setR[\numplus]}} \quad&{\sum_{\idxplus=1}^{\numplus} \U^{-1}\paren*{\yplus[\idxplus]}\quad\paren*{\text{\scriptsize$=\sum_{\idxplus=1}^{\numplus}\paren*{\yplus}^{\frac{1}{\alpha}}$}}}\\
            \subjectto\quad&0\le \yplus[1]\leq \yplus[2] \leq \cdots \leq \yplus[\numplus],\\
            \quad&\sum_{\idxplus=1}^{\numplus}\hplus\yplus=v.
        \end{split}
    \end{align}
    \end{proposition}
    \begin{proof}
        By changing the variables from $\paren*{\wminus, \wplus[]}$ to $\paren*{\yminus[], \yplus[]}$, the problem 
        \cref{prob:orgmaxprofit}
        reduces to the following form:
        \begin{subequations}\label{prob:yorgopt}
            \begin{align}
                \min_{\substack{\yminus[]\in\setR[\numminus],\,\yplus[]\in\setR[\numplus]}}\quad&{\sum_{\idxminus=1}^{\numminus} \U^{-1}\paren*{-\yminus[\idxminus]} + \sum_{\idxplus=1}^{\numplus} \U^{-1}\paren*{\yplus[\idxplus]} \protect\label{objfun:CPTprofit}}\\
                \subjectto\quad&{
                - \sum_{\idxminus=1}^{\numminus}\hminus \yminus[\idxminus] + \sum_{\idxplus=1}^{\numplus}\hplus \yplus[\idxplus] \ge 0, \protect\label{cstr:yCPTexpectedutility}}\\
                \quad&{0\leq \yplus[1] \leq \cdots \leq \yplus[\numplus] \protect\label{cstr:ygainascending}},\\
                \quad&{0\leq \yminus[1] \leq \cdots \leq \yminus[\numminus] \protect\label{cstr:ylossascending}},
            \end{align}
        \end{subequations}
        where \cref{cstr:ygainascending,cstr:ylossascending} follow from \cref{cstr:gainascending,cstr:lossascending} and the strict monotonicity of the value function \cref{def:utilityfunc}.
        Then, we introduce a new variable $\eqval \in \setR_{\geq 0}$ and add a constraint $\eqval = \sum_{\idxplus=1}^{\numplus}\hplus \yplus[\idxplus]$ to the problem \cref{prob:yorgopt}. 
        
        Since the objective function~\cref{objfun:CPTprofit} is monotone decreasing with respect to $\yminus[1],\ldots,\yminus[\numminus]$, 
        we can reduce \cref{cstr:yCPTexpectedutility} to the equality constraints
        $\sum_{\idxminus=1}^{\numminus}\hminus \yminus[\idxminus] = \sum_{\idxplus=1}^{\numplus}\hplus \yplus[\idxplus] \paren*{=\eqval}.$
        Thus, under a fixed $\eqval$, we can independently optimize $\yminus[]$ and $\yplus[]$.
        Then, by optimizing $\eqval \in \setR_{\geq 0}$, we obtain the entire solution to 
        \cref{prob:yorgopt},
        which leads to the equivalent formulation \cref{prob:middlevsubprob,cstr:4thlosssubprob,cstr:4thgainsubprob}. 
    \end{proof}
       
    In summary, we formulate the design of the optimal lottery as a three-level optimization problem; the high-level problem is \cref{prob:orgNminusplus}, where we optimize the number of gains and losses to maximize the seller's profit.
    The middle-level problem is \cref{prob:middlevsubprob}, which is equivalent to the problem 
    \cref{prob:orgmaxprofit}. 
    The problem \cref{prob:middlevsubprob} is devoted to calculating the optimal profit of the seller under the fixed number of gains and losses.
    To solve \cref{prob:middlevsubprob}, we introduce 
    two low-level problems \cref{cstr:4thlosssubprob,cstr:4thgainsubprob},
    where we essentially compute the optimal design of lotteries under the fixed number of gains and losses, respectively.
    Let $\paren*{\numminus^*,\numplus^*}$ be the optimal solution of \cref{prob:orgNminusplus} and 
    let $\paren*{\wminus[]^*,\wplus[]^*}\in\setR[\numminus^*]\times\setR[\numplus^*]$ be the optimal solution of 
    \cref{prob:orgmaxprofit}.
    Then, an optimal ticket price is $t^*=-\min_{i=1}^{\numminus^*}\wminus[i]^*$ and optimal lottery prizes are $\paren*{\wminus[1]^*-t^*,\dots,\wminus[\numminus^*]^*-t^*,\wplus[1]^*-t^*,\dots,\wplus[\numplus^*]^*-t^*}$.
    In the next section, we solve the above problems 
    \cref{cstr:4thlosssubprob,cstr:4thgainsubprob,prob:middlevsubprob} one by one.
    \begin{remark}
        We note that our formulation directly covers simpler settings;
        by setting $W\paren*{p} = \weightfuncminus\paren*{p} = p$ ($\weightfuncparam=\weightfuncparamminus=1$), our formulation reduces to the EU framework.
        Hence, our analysis includes the optimal lottery design based on EU.
        We can also deal with a setting where the value function is omitted (i.e., $V\paren*{\omega} = \omega$) by taking $\alpha=\beta=1$ in \cref{prob:middlevsubprob}.
        Although we assume $\alpha, \beta\in\paren*{0,1}$ in this paper, it is not difficult to check that our analysis can be easily extended to $\alpha=1$ or $\beta=1$.
    \end{remark}
    
    \begin{remark}
        In the case of RDEU, where the value function is globally concave, the problem 
        \cref{prob:orgmaxprofit}
        reduces to a convex problem. Thus, we can efficiently calculate its optimal solution using a standard convex problem solver.
        However, in the case of CPT, the problem 
        \cref{prob:orgmaxprofit}
        is nonconvex; therefore, such a solver is not applicable. 
    \end{remark}

\section{Analysis and algorithms}
\label{sec:analysis}
    In this section, we propose a linear-time algorithm to solve the three-level optimization problem consisting of the high-level one~\cref{prob:orgNminusplus}, the middle-level one \cref{prob:middlevsubprob}, and the two low-level ones
    \cref{cstr:4thlosssubprob,cstr:4thgainsubprob}.
    To this end, we provide the algorithmic solutions of the problems~\cref{cstr:4thlosssubprob,cstr:4thgainsubprob} 
    in \cref{subsec:lowlevelgainanalysis,subsec:lowlevellossanalysis}, respectively.
    In \cref{subsec:middlelevelanalysis}, we derive the analytic solution of \cref{prob:middlevsubprob}.
    In \cref{subsec:highlevelanalysis}, we propose a linear-time algorithm to solve \cref{prob:orgNminusplus} by combining the solutions above.
    For notational simplicity, we define $\balpha \coloneqq 1/\alpha$ and $\bbeta \coloneqq 1/\beta$.
    Note that, by the assumptions of $\alpha,\beta\in \paren*{0,1}$, we have $\balpha,\bbeta > 1$.
        
\subsection{\texorpdfstring{Analysis of low-level problem on gains \cref{cstr:4thgainsubprob}}{Analysis of low-level problem on gains} \label{subsec:lowlevelgainanalysis}}
    In this subsection, we provide the optimal solution to the problem 
    \cref{cstr:4thgainsubprob}.
    The problem 
    \cref{cstr:4thgainsubprob} itself is convex; hence, its optimal solution can be numerically obtained by using a solver. 
    However, our analysis provides the optimal solution in an analytical formula, which is crucial for considering a linear-time algorithm. 
    The optimal value is clearly zero if $\eqval=0$. 
    Thus, we only consider the case $\eqval>0$ in the following.
    If $\numplus=1$, the feasible region is a singleton $\brc*{\eqval/\hplus[1]}$.
    Otherwise (i.e., $\numplus>1$), the feasible region has an interior point, for example, $\yplus= 2\idxplus\eqval\cdot \sbra*{\numplus\paren*{\numplus+1}\sum_{\idxplus=1}^\numplus\hplus}^{-1}$ 
    for $\idxplus\in\setIntvl[\numplus]$.
    Since the objective function is convex and the constraints are linear, the problem 
    \cref{cstr:4thgainsubprob}
    is a convex optimization problem.
    Therefore, the Karush--Kuhn--Tucker (KKT) conditions are necessary and sufficient for the optimality (see, e.g., \cite{BoydVandenberghe04ConvexOptBook}).
            
    To start with, we define a key index for our analysis.
    \begin{definition}\label{defi:transididx}
    The transitional index $\transidplus\in\setIntvl[\numplus]$ is defined as
    \begin{align}\label{eq:transiddef}
        \transidplus \coloneqq \min\brc*{\idxplus \in \setIntvl[\numplus] \relmiddle{|} \idxplus\cdot \hplus[\idxplus+1]\geq \sum_{\idxplus'=1}^{\idxplus}\hplus[\idxplus']},
    \end{align}
    where we treat $\hplus[\numplus+1]$ as $\infty$.
    \end{definition}
    In words, $\transidplus$ is the index at which $\hplus[\transidplus+1]$ is equal to or higher than the previous average.
    Note that since $\hplus$ only depends on $\numplus$ for every $\idxplus\in\setIntvl[\numplus]$, the transitional index $\transidplus$ can be parameterized by $\numplus$ if we regard $\numplus$ as a variable. 
    In \cref{subsec:middlelevelanalysis,subsec:highlevelanalysis}. we will use $\transidplus[\numplus]$ for the transitional index instead of $\transidplus$ to denote this parameterization. 
    Now, we are ready to give the optimal solution of the problem 
    \cref{cstr:4thgainsubprob}.
    \begin{theorem}\label{THEO:PLUS}
        The optimal solution of
        the problem 
        \cref{cstr:4thgainsubprob} 
        is $\yoptplus[]\in\setR[\numplus]$ such that
        \begin{align}\label{eq:thmyoptplus}
            \yoptplus =
            \begin{cases}
                \optconstplus & \paren*{1\leq \idxplus\leq \transidplus},\\
                \paren*{\frac{J\hplus}{\sum_{\idxplus'=1}^{\transidplus}\hplus[\idxplus']}}^{\frac{\alpha}{1-\alpha}} \optconstplus & \paren*{\transidplus+1\leq \idxplus\leq \numplus}
            \end{cases}
        \end{align}
        for all $\idxplus\in\setIntvl[\numplus]$,
        where the constant $\optconstplus\geq 0$ is defined as
        \begin{align*}
            \optconstplus \coloneqq \frac{v\paren*{\sum_{\idxplusprime=1}^{\transidplus}\hplus[\idxplusprime]}^{\frac{\alpha}{1-\alpha}}}{\paren*{\sum_{\idxplusprime=1}^{\transidplus}\hplus[\idxplusprime]}^{\frac{1}{1-\alpha}}+\transidplus^{\frac{\alpha}{1-\alpha}}\sum_{\idxplusprime=\transidplus+1}^{\numplus}\hplus[\idxplusprime]^{\frac{1}{1-\alpha}}}.
        \end{align*}
    \end{theorem}
    \cref{THEO:PLUS} shows that, under the fixed number of gain tickets $\numplus$, the optimal solution consists of uniform prizes and increasing ones.
    From this theorem, the optimal value of the problem 
    \cref{cstr:4thgainsubprob}
    is given as follows.
    \begin{corollary}
        \label{corolally:posioptval}
        The optimal value of the problem 
        \cref{cstr:4thgainsubprob} is
        \begin{align*}
            \eqval^{\frac{1}{\alpha}}\cdot \transidplus\cdot\sbra*{\paren*{\sum_{\idxplus=1}^{\transidplus} \hplus}^\frac{1}{1-\alpha}+\sum_{\idxplus=\transidplus+1}^\numplus\paren*{{\transidplus}^{\alpha}\hplus}^\frac{1}{1-\alpha}}^{-\frac{1-\alpha}{\alpha}}.
        \end{align*}
    \end{corollary}
    In the rest of this subsection, we prove \cref{THEO:PLUS} by explicitly identifying the corresponding Lagrange multipliers and confirming that they satisfy the KKT conditions.
    To this end, we first introduce an additional definition, the \textit{peak index}.
    We will use this index to divide the behaviors of $\yoptplus$ and the multipliers into several cases.
    \begin{definition}[peak index]
        \label{defi:peakindex}
        We say that $\flexplus \in \setIntvl[\numplus]$ is a peak index if it satisfies that
        $\hplus[\numplus] > \hplus[\numplus-1] > \dots > \hplus[\flexplus] \text{ and } \hplus[\flexplus] \leq \hplus[\flexplus-1] < \hplus[\flexplus-2] < \dots < \hplus[1].$
    \end{definition}
    The following lemma ensures the existence of a peak index.
    \begin{lemma}\label{lemm:existflexind}
        Suppose that the probability weighting function $\weightfuncplus\colon\sbra*{0,1}\to\sbra*{0,1}$ is inverse S-shaped. Then, there exists a unique peak index $\flexplus \in \setIntvl[\numplus]$.
    \end{lemma}
    \begin{proof}
        Since $\weightfuncplus\paren*{0}=0$ holds, it follows that
        \begin{align}\label{eq:hplusWplusminusWplus}
            \hplus=\weightfuncplus\paren*{\frac{\numplus-\idxplus+1}{\N}}-\weightfuncplus\paren*{\frac{\numplus-\idxplus}{\N}}
        \end{align}
        for every $\idxplus \in \setIntvl[\numplus]$. 
        Then, for all $\idxplus \in \setIntvl[\numplus]$, applying the mean value theorem to the left-hand side in \cref{eq:hplusWplusminusWplus} implies that there exists 
        \begin{align*}
            \theta_{\idxplus} \in \paren*{\frac{\numplus-\idxplus}{\N}, \frac{\numplus-\idxplus+1}{\N}} \text{ such that }\hplus=\frac{\dv{\weightfuncplus}{p}\paren*{\theta_{\idxplus}}}{\N}.
        \end{align*}
        Let $x_{0}$ be the inflection point of $\weightfuncplus$ and $\tildeidxplus$ be the minimum index satisfying $\theta_{\tildeidxplus} \leq x_0$ ($\tildeidxplus=1$ if $\theta_1>x_0$).
        Since $\weightfuncplus$ is an inverse S-shaped function, it follows that
        $\hplus[\numplus] > \hplus[\numplus-1] > \dots > \hplus[\tildeidxplus] \text{ and } \hplus[\tildeidxplus-1] < \hplus[\tildeidxplus-2] < \dots < \hplus[1]$.
        Then
        \begin{align*}
            \flexplus=\begin{cases}
            \tildeidxplus    & \paren*{\text{if }\tildeidxplus=1 \text{ or }\hplus[\tildeidxplus]\leq\hplus[\tildeidxplus-1]},\\
            \tildeidxplus -1 & \paren*{\text{otherwise}}
            \end{cases}
        \end{align*}
        is the unique peak index.
    \end{proof}
    
     We note that by the inverse S-shape of the weighting function, it follows from \cref{defi:peakindex} that $\transidplus\geq\flexplus$. Especially, we have $\transidplus=1$ if and only if $\flexplus=1$. 
    
    Next, we give the optimality condition of the problem 
    \cref{cstr:4thgainsubprob}.
    As we mentioned, it is sufficient to consider the KKT conditions. The Lagrangian of the problem 
    \cref{cstr:4thgainsubprob}
    is given by
    \begin{align*}
      L\paren*{\yplus[],\mu,\Lambda} \coloneqq \sum_{\idxplus=1}^\numplus\paren*{\yplus}^\balpha -\mu_1\yplus[1] +\sum_{\idxplus=1}^{\numplus-1}\mu_{\idxplus+1}\paren*{\yplus-\yplus[\idxplus+1]}+\Lambda\paren*{\sum_{\idxplus=1}^\numplus\hplus\yplus-\eqval},
    \end{align*}
    where $\lagrangemu[] = \paren*{\lagrangemu[1],\dots,\lagrangemu[\numplus]}\in \setR[\numplus]$ and $\lagrangelambda\in \setR[]$ are Lagrange multipliers. Then, the KKT conditions are 
    \begin{alignat}{2}
        \label{KKT:stat1}
        \balpha \paren*{\yplus}^{\balpha-1} +\mu_{\idxplus+1}-\mu_\idxplus+\Lambda \hplus &= 0 \quad&& \paren*{\idxplus\in\setIntvl[\numplus-1]},\\
        \label{KKT:stat2}
        \balpha \paren*{\yplus[\numplus]}^{\balpha-1} -\mu_\numplus+\Lambda \hplus[\numplus] &=0,&&\\
        \label{KKT:comp}
        \mu_1\yplus[1]= 0,~\mu_{\idxplus+1}\paren*{\yplus-\yplus[\idxplus+1]} &= 0 && \paren*{\idxplus\in\setIntvl[\numplus-1]},\\
        \label{KKT:equal}
        \sum_{\idxplus=1}^\numplus\hplus\yplus&=\eqval,&&\\
        \label{KKT:ineq1}
        0\leq \yplus[1]\leq \dots &\leq \yplus[\numplus],\\
        \label{KKT:ineq2}
        \lagrangemu&\geq 0 && \paren*{ \idxplus\in\setIntvl[\numplus]},
    \end{alignat}
    where \cref{KKT:stat1} and \cref{KKT:stat2} are obtained from $\nabla_{\yplus[]} L=0$, \cref{KKT:comp} is the complementary condition, and \cref{KKT:equal}, \cref{KKT:ineq1} and \cref{KKT:ineq2} are by the condition for the Lagrange multipliers with respect to inequality constraints and the constraints of the original problem.
    
    Then, we give Lagrange multipliers satisfying \cref{KKT:stat1}--\cref{KKT:ineq2} with $\yoptplus$ defined in \cref{THEO:PLUS}.
    
    \begin{lemma}
        \label{lemm:KKT}
        Define $\yoptplus[]\in\setR[\numplus]$ as in \cref{THEO:PLUS} and
        \begin{align}
            \label{equation:mu}
                \optlagrangemu[\idxplus] =& -\frac{\balpha}{v}\sbra*{\sum_{\idxplusprime=1}^\numplus\paren*{\yoptplus[\idxplusprime]}^\balpha}\paren*{\sum_{\idxplusprime=\idxplus}^\numplus\hplus[\idxplusprime]}+\balpha\sbra*{\sum_{\idxplusprime=\idxplus}^\numplus\paren*{\yoptplus[\idxplusprime]}^{\balpha-1}}~\paren*{\idxplus\in\setIntvl[n]},\\
            \optlagrangelambda =& -\frac{\balpha\sum_{\idxplusprime=1}^\numplus\paren*{\yoptplus[\idxplusprime]}^\balpha}{\eqval}.\notag
        \end{align}
        Then, $\paren*{\paren*{\yoptplus[\idxplus]}_{\idxplus=1}^\numplus, \paren*{\optlagrangemu}_{\idxplus=1}^\numplus, \optlagrangelambda}$
        satisfy the KKT conditions~\cref{KKT:stat1}--\cref{KKT:ineq2}.
    \end{lemma}

    \begin{proof}
        The equalities \cref{KKT:stat1,KKT:stat2,KKT:equal} can be checked by straightforward calculations.
        Thus, it is sufficient to confirm \cref{KKT:comp,KKT:ineq1,KKT:ineq2}.
        Since $\transidplus\geq \flexplus$, we can observe that $\paren*{\yoptplus}_{\idxplus=1}^\numplus$ in \cref{THEO:PLUS} satisfies
            $0\leq \yoptplus[1]\leq \dots\leq \yoptplus[\numplus]$,
        that is, the inequalities \cref{KKT:ineq1} hold. 
        Moreover, we have $\yoptplus-\yoptplus[\idxplus+1]=0$ for every $\idxplus\in\setIntvl[\transidplus-1]$. Therefore, it suffices to check the $\paren*{\optlagrangemu}_{\idxplus=1}^\numplus$ determined by \cref{equation:mu} satisfies complementary condition \cref{KKT:comp} and the inequalities \cref{KKT:ineq2}:
        $\optlagrangemu[\idxplus]\geq 0$ for all $\idxplus\in\setIntvl[\numplus]$ and $\optlagrangemu[\idxplus]=0$ for $\idxplus=1$ and $\transidplus+1\leq \idxplus\leq n$.

        Define $t_\idxplus \coloneqq \paren*{\yoptplus}^{\balpha-1}/\hplus$ for all $\idxplus\in\setIntvl[\numplus]$. 
        By the construction of $\yoptplus[]$, a straightforward calculation shows $\sum_{j=1}^\numplus\hplus\yoptplus = \eqval$.
        Thus, by \cref{equation:mu}, we have
        \begin{align}
            \begin{split}\label{equation:mun}
                -\frac{\eqval}{\balpha}\optlagrangemu[\numplus] &= \hplus[\numplus]\sum_{\idxplusprime=1}^\numplus\paren*{\yoptplus[\idxplusprime]}^\balpha-\eqval\paren*{\yoptplus[\numplus]}^{\balpha-1}
                =\hplus[\numplus]\sum_{\idxplusprime=1}^\numplus\paren*{\yoptplus[\idxplusprime]}^\balpha-\paren*{\sum_{\idxplusprime=1}^\numplus\hplus[\idxplusprime]\yoptplus[\idxplusprime]}\paren*{\yoptplus[\numplus]}^{\balpha-1}\\
                &=\sum_{\idxplusprime=1}^\numplus\yoptplus[\idxplusprime]\sbra*{\hplus[\numplus]\paren*{\yoptplus[\idxplusprime]}^{\balpha-1}-\hplus[\idxplusprime]\paren*{\yoptplus[\numplus]}^{\balpha-1}} 
                =\sum_{\idxplusprime=1}^\numplus\hplus[\idxplusprime]\hplus[\numplus]\yoptplus\paren*{t_\idxplusprime-t_\numplus},
            \end{split}
        \end{align}
        and  
        \begin{align}
            \begin{split}\label{equation:mudiff}
                \frac{\eqval}{\balpha}\paren*{\optlagrangemu[\idxplus+1]-\optlagrangemu[\idxplus]} &= \hplus\sum_{\idxplusprime=1}^\numplus\paren*{\yoptplus[\idxplusprime]}^\balpha-\eqval\paren*{\yoptplus}^{\balpha-1}
                =\hplus\sum_{\idxplusprime=1}^\numplus\paren*{\yoptplus[\idxplusprime]}^\balpha-\paren*{\sum_{j=1}^\numplus\hplus[\idxplusprime]\yoptplus[\idxplusprime]}\paren*{\yoptplus}^{\balpha-1}\\
                &=\sum_{\idxplusprime=1}^\numplus\yoptplus[\idxplusprime]\sbra*{\hplus\paren*{\yoptplus[\idxplusprime]}^{\balpha-1}-\hplus[\idxplusprime]\paren*{\yoptplus}^{\balpha-1}}
                =\sum_{\idxplusprime=1}^\numplus\hplus[\idxplusprime]\hplus\yoptplus[\idxplusprime]\paren*{t_\idxplusprime-t_\idxplus}
            \end{split}
        \end{align}
        for each $\idxplus\in\setIntvl[\numplus-1]$.

        In what follows, we separately consider two cases: (i) $\transidplus<\numplus$ and (ii) $\transidplus=\numplus$.
        
        \paragraph{Case (i)} 
        Suppose that $\transidplus<\numplus$. 
        In this case, we have
        \begin{align*}
        t_{\transidplus+1}=\dots=t_\numplus=\frac{\transidplus \optconstplus[\balpha-1]}{\sum_{\idxplus'=1}^{\transidplus}\hplus[\idxplus']} \eqqcolon T.
        \end{align*}
        Then, by \cref{equation:mun}, we obtain
        \begin{align*}
            -\frac{\eqval}{\balpha}\optlagrangemu[\numplus] &= \sum_{\idxplusprime=1}^\numplus\hplus[\idxplusprime]\hplus[\numplus]\yoptplus[\idxplusprime]\paren*{t_\idxplusprime-t_\numplus}
            = \hplus[\numplus]\sum_{\idxplusprime=1}^{\transidplus}\hplus[\idxplusprime] \optconstplus[]\paren*{t_\idxplusprime-t_\numplus}
            = \hplus[\numplus]\sum_{\idxplusprime=1}^{\transidplus}\paren*{\optconstplus[\balpha]-\frac{\transidplus\hplus[\idxplusprime] }{\sum_{\idxplus'=1}^{\transidplus}\hplus[\idxplus']}\optconstplus[\balpha]}
            = \hplus[\numplus]\paren*{\transidplus\optconstplus[\balpha]-\transidplus\optconstplus[\balpha]}=0.
        \end{align*}
        We can also get $\optlagrangemu[\idxplus+1]-\optlagrangemu[\idxplus]=0$ for $\idxplus=\transidplus+1,\dots,\numplus-1$ by replacing $t_\numplus$ by $t_\idxplus$ in the above transformation.
        Hence, we get $\optlagrangemu[\idxplus]=0$ for $\transidplus+1\leq \idxplus\leq \numplus$.
        
        Moreover, \cref{equation:mu} implies
        \begin{align}
        -\frac{\eqval}{\balpha}\optlagrangemu[1] &=
        -\frac{\eqval}{\balpha}\optlagrangemu[\numplus]+\frac{\eqval}{\balpha}\sum_{\idxplus=1}^{\numplus-1}\paren*{\optlagrangemu[\idxplus+1]-\optlagrangemu[\idxplus]}\notag
        \notag
            = \sum_{\idxplus_1,\idxplus_2=1}^\numplus\hplus[\idxplus_1]\hplus[\idxplus_2]\yoptplus[\idxplus_1]\paren*{t_{\idxplus_1}-t_{\idxplus_2}}\notag\\
           \label{eq:mu1calc} &=\sum_{\idxplus_1<\idxplus_2}\hplus[\idxplus_1]\hplus[\idxplus_2]\paren*{\yoptplus[\idxplus_1]-\yoptplus[\idxplus_2]}\paren*{t_{\idxplus_1}-t_{\idxplus_2}}\\
            &= \sum_{\idxplus_1=1}^{\transidplus} \sum_{\idxplus_2=\transidplus+1}^{\numplus} \hplus[\idxplus_1]\hplus[\idxplus_2]\paren*{\yoptplus[\idxplus_2]-\yoptplus[\idxplus_1]}\paren*{t_{\idxplus_2}-t_{\idxplus_1}}\notag\\
            &= \sum_{\idxplus_1=1}^{\transidplus} \sum_{\idxplus_2=\transidplus+1}^{\numplus} \hplus[\idxplus_1]\hplus[\idxplus_2]\paren*{\yoptplus[\idxplus_2]-Y}\paren*{T-t_{\idxplus_1}}\notag\\
            &=\sum_{\idxplus_1=1}^{\transidplus}\hplus[\idxplus_1]\paren*{T-t_{\idxplus_1}}\sum_{\idxplus_2=\transidplus+1}^{\numplus}\hplus[\idxplus_2]\paren*{\yoptplus[\idxplus_2]-Y}\notag\\
            &= \paren*{\transidplus[]\optconstplus[\balpha]-\transidplus[]\optconstplus[\balpha]}\sum_{\idxplus_2=\transidplus+1}^{\numplus}\hplus[\idxplus_2]\paren*{\yoptplus[\idxplus_2]-Y} = 0,\notag
        \end{align}
        where the second equality follows from \cref{equation:mun} and \cref{equation:mudiff}, the
        fourth equality from $\yoptplus[\idxplus_1]=\yoptplus[\idxplus_2]$ when $\idxplus_2\leq \transidplus$ and $t_{\idxplus_1}=t_{\idxplus_2}$ when $\idxplus_1\geq \transidplus+1$, and the fifth equality from the definition of $T$ and $\hplus t_{\idxplus} = \paren*{\yoptplus}^{\balpha-1}=Y^{\balpha-1}$ for $\idxplus \in \setIntvl[\transidplus]$. 
        Now, it remains to prove that $\optlagrangemu[\idxplus]\geq 0$ for $1< \idxplus\leq \transidplus$. 
        For $\idxplus \in \setIntvl[\transidplus]$, we have
        \begin{align}
            \label{eq:toy}
            \optlagrangemu[\idxplus+1]-\optlagrangemu[\idxplus] &=\balpha\sbra*{\frac{\hplus[\idxplus]}{\eqval}\sum_{\idxplusprime=1}^\numplus\paren*{\yoptplus[\idxplusprime]}^\balpha-\paren*{\yoptplus[\idxplus]}^{\balpha-1}}=\balpha\sbra*{\frac{\hplus}{\eqval}\sum_{\idxplusprime=1}^\numplus\paren*{\yoptplus[\idxplusprime]}^\balpha-\optconstplus[\balpha-1]}.
        \end{align}
        Since $\paren*{\transidplus-1}\hplus[\transidplus] < \sum_{\idxplusprime=1}^{\transidplus-1}\hplus[\idxplusprime]$ holds by the definition of $\transidplus$, we have
        \begin{align*}
            t_{\transidplus} = \frac{\optconstplus[\balpha-1]}{\hplus[\transidplus]} > \frac{\transidplus\optconstplus[\balpha-1]}{\sum_{\idxplusprime=1}^{\transidplus}\hplus[\idxplusprime]}=\tconst.
        \end{align*}
        This implies 
        \begin{align*}
            \frac{\eqval}{\balpha}\paren*{\optlagrangemu[\transidplus+1]-\optlagrangemu[\transidplus]}                &=\sum_{\idxplusprime=1}^{\numplus}\hplus[\idxplusprime]\hplus[\transidplus]\yoptplus[\idxplusprime]\paren*{t_\idxplusprime-t_{\transidplus}} \cdot 
            \sum_{\idxplusprime=1}^{\numplus}\hplus[\idxplusprime]\hplus[\transidplus]\yoptplus[\idxplusprime]\paren*{t_{\idxplusprime}-T}
            = -\frac{\hplus[\transidplus]}{\hplus[\numplus]} \cdot \frac{\eqval}{\balpha}\optlagrangemu[\numplus]=0.
        \end{align*}
        Combining this with $\optlagrangemu[\transidplus+1]=0$, we get $\optlagrangemu[\transidplus]>0$.
        In addition, since the  increase/decrease of the right-hand side in \cref{eq:toy} is determined by that of $\hplus$, the definition of $\flexplus$ gives
        \begin{equation}
            \optlagrangemu[2]-\optlagrangemu[1] > \dots >\optlagrangemu[\flexplus]-\optlagrangemu[\flexplus-1] \geq \optlagrangemu[\flexplus+1]-\optlagrangemu[\flexplus]
            \label{eq:mudiffineq1}
        \end{equation}
        and
        \begin{align}
            \label{eq:mudiffineq2}
            \optlagrangemu[\flexplus+1]-\optlagrangemu[\flexplus] < \dots < \optlagrangemu[\transidplus+1] - \optlagrangemu[\transidplus]<0.
        \end{align}
        Then, we can show that there exists an integer $\transidplus_0\in\sbra*{\transidplus-1}$ such that
        \begin{align}
        \label{eq:mudiffposinega}
            \begin{cases} 
                \optlagrangemu[\idxplus+1]-\optlagrangemu[\idxplus] \geq 0 &\paren*{1\le\idxplus \le \transidplus[0]},\\
                \optlagrangemu[\idxplus+1]-\optlagrangemu[\idxplus] < 0 &\paren*{\transidplus_0\le \idxplus \leq \transidplus}.
            \end{cases}
        \end{align}
        Indeed, if there are no such indices, it holds that $\optlagrangemu[\idxplus+1]-\optlagrangemu[\idxplus]<0$ for all $\idxplus\in\setIntvl[\transidplus]$ by \cref{eq:mudiffineq1} and \cref{eq:mudiffineq2}. This implies that $\paren*{\optlagrangemu[\idxplus]}_{\idxplus=1}^{\transidplus+1}$ is monotonically increasing, which contradicts to $\optlagrangemu[1]=\optlagrangemu[\transidplus+1]=0$.
        \cref{eq:mudiffposinega} induces
        \begin{align*}
            0 = \optlagrangemu[1]\leq \optlagrangemu[2]\leq \dots\leq \optlagrangemu[\transidplus_{0}+1]
            \text{ and }
            \optlagrangemu[\transidplus_{0}+1]>\optlagrangemu[\transidplus_{0} + 2]> \dots>\optlagrangemu[\transidplus+1] = 0.
        \end{align*}
        Thus, we get $\optlagrangemu[\idxplus]\geq 0$ for $1< \idxplus\leq \transidplus$.
        Hence, \cref{KKT:comp} and \cref{KKT:ineq2} hold, which gives the conclusion for the case $\transidplus<\numplus$.

        \paragraph{Case (ii)}
        Suppose that $\transidplus=\numplus$. 
        In this case, \cref{eq:thmyoptplus} implies $\yoptplus=\optconstplus[]$ for all $\idxplus\in\setIntvl[\numminus]$. Thus, it suffices to show that $\optlagrangemu\geq 0$ for all $\idxplus\in\setIntvl[\numplus]$. We can get $\optlagrangemu[1]=0$ by
        \begin{align*}
            -\frac{\eqval}{\balpha}\optlagrangemu[\idxplus] &= \sum_{\idxplus_1=1}^{\transidplus} \sum_{\idxplus_2=\transidplus+1}^{\numplus} \hplus[\idxplus_1]\hplus[\idxplus_2]\paren*{\yoptplus[\idxplus_2]-\yoptplus[\idxplus_1]}\paren*{t_{\idxplus_2}-t_{\idxplus_1}}
            =\sum_{\idxplus_1=1}^{\transidplus} \sum_{\idxplus_2=\transidplus+1}^{\numplus} \hplus[\idxplus_1]\hplus[\idxplus_2]\paren*{\optconstplus-\optconstplus}\paren*{t_{\idxplus_2}-t_{\idxplus_1}}=0,
        \end{align*}
        where the fist equality is the same as \cref{eq:mu1calc}.
        For $\idxplus\geq 2$, \cref{equation:mu} gives
        \begin{align*}
            \frac{\eqval}{\balpha}\optlagrangemu[\idxplus]
            &= -\sbra*{\sum_{\idxplusprime=1}^\numplus\paren*{\yoptplus[\idxplusprime]}^\balpha}\paren*{\sum_{\idxplusprime=\idxplus}^\numplus\hplus[\idxplusprime]}+\balpha\sbra*{\sum_{\idxplusprime=\idxplus}^\numplus\paren*{\yoptplus[\idxplusprime]}^{\balpha-1}}
            = -\numplus \optconstplus[\balpha]\paren*{\sum_{\idxplus'=1}^\idxplus\hplus[\idxplus']}+\eqval\paren*{\numplus-\idxplus+1}\optconstplus[\balpha-1]\\
            &=Y^{\balpha-1}\sbra*{\eqval\paren*{\numplus-\idxplus+1}-\numplus \paren*{\sum_{\idxplus'=\idxplus}^{\numplus}\hplus[\idxplus'] \optconstplus[]}}
            =Y^{\balpha-1}\sbra*{\eqval\paren*{\numplus-\idxplus+1}-\numplus \paren*{v-\sum_{\idxplus'=1}^{\idxplus-1}\hplus[\idxplus'] Y}}\\
            &=\paren*{\idxplus-1}Y^{\balpha-1}\paren*{\frac{\sum_{\idxplus'=1}^{\idxplus-1}\hplus[\idxplus']}{\idxplus-1}\optconstplus\numplus-\eqval} \geq \paren*{\idxplus-1}Y^{\balpha-1}\paren*{\frac{\sum_{\idxplus'=1}^{\numplus}\hplus[\idxplus']}{\numplus}\optconstplus[]\numplus-\eqval}
            =\paren*{\idxplus-1}\optconstplus[\balpha-1]\paren*{\eqval-\eqval}=0,
        \end{align*}
        where the last equality follows from \cref{KKT:equal}.
        The inequality follows from the fact that the average of $\hplus[\idxplusprime]$, that is, $\paren*{\sum_{\idxplus'=1}^\idxplus\hplus[\idxplus']/\idxplus}_{\idxplus=1}^{\numplus}$, is monotonically non-increasing by the definition of $\transidplus$. 
        Thus we get $\optlagrangemu\geq 0$ for all $\idxplus\in\setIntvl[\numplus]$, which gives the conclusion for the case $\transidplus=\numplus$. 
    \end{proof}

    Now, we are ready to prove the optimality of $\yoptplus[]$.
    
    \begin{proof}[Proof of \cref{THEO:PLUS}]
        \cref{lemm:KKT} shows that $\yoptplus[]$ defined in \cref{THEO:PLUS} is the KKT point of the problem 
        \cref{cstr:4thgainsubprob}. 
        Since the problem 
        \cref{cstr:4thgainsubprob}
        is the convex optimization problem and its objective function is strongly convex, $\yoptplus[]$ is a unique optimum.
    \end{proof}

\subsection{Analysis of low-level problem on losses 
    \texorpdfstring{\cref{cstr:4thlosssubprob}}{\space}}
    \label{subsec:lowlevellossanalysis}
    In this subsection, we provide the optimal solution to the problem 
    \cref{cstr:4thlosssubprob}.
    Note that the problem 
    \cref{cstr:4thlosssubprob}
    is a nonconvex optimization problem.
    However, it has a global optimum because the objective function is continuous, and the feasible region is a compact subset of $\setR[\numminus]$.
    Moreover, by utilizing the structure of the feasible region and the concavity of the objective function, we can construct the optimal solution of the problem 
    \cref{cstr:4thlosssubprob}
    as follows.
    \begin{theorem}
    \label{prop:minusoptimum}
        There exists an index $\lminus\in\setIntvl[\numminus]$ such that
        \begin{align}
            \label{eq:vartex}
            \yoptminus =
            \begin{cases}
                \optconstminus & \paren*{1 \leq \idxminus\leq \lminus},\\
                0 & \paren*{\lminus+1\leq \idxminus\leq \numminus}
            \end{cases}
        \end{align}
        is a global optimum of the problem 
        \cref{cstr:4thlosssubprob}, 
        where $\optconstminus \coloneqq \eqval/\sum_{\idxminus=1}^{\lminus}\hminus~\paren*{\ge 0}$.
    \end{theorem}
    
    \cref{prop:minusoptimum} shows that, under the fixed number of loss tickets $\numminus$, the optimal solution consists of lotteries with no prize or a payment equal to the ticket price. In \cref{subsec:highlevelanalysis}, we reveal that the latter part does not need to be considered 
    in the optimum solution of the high-level problem \cref{prob:orgNminusplus}.
    The following corollary specifies the optimal value of the problem 
    \cref{cstr:4thlosssubprob}.
    
    \begin{corollary}
        \label{theo:minus}
        The optimal value of the problem 
        \cref{cstr:4thlosssubprob}
        is 
        \begin{align}\label{eq:lowminusoptval}
        -\paren*{\frac{\eqval}{\propminus}}^{\frac{1}{\beta}} \cdot \max_{\lminus\in\sbra*{\numminus}} \lminus\cdot\paren*{\sum_{\idxminus=1}^{\lminus}\hminus}^{-\frac{1}{\beta}}.
        \end{align}
    \end{corollary}
    Note that, since $\hminus$ only depends on $\numminus$ for every $\idxminus\in\setIntvl[\numminus]$, the index that realizes \cref{eq:lowminusoptval} can be parameterized by $\numminus$ if we regard $\numminus$ as a variable.
    In \cref{subsec:middlelevelanalysis,subsec:highlevelanalysis}. we will use the symbol $\lminus[\numminus]$ instead of $\lminus$ to denote the parameterization; that is, we define
    \begin{align}\label{def:lminusnumminus}
        \lminus[\numminus] \coloneqq \arg\max_{\lminus\in\setIntvl[\numminus]} \, \lminus\cdot\paren*{\sum_{\idxminus=1}^{\lminus}\hminus}^{-\frac{1}{\beta}}.
    \end{align}
    If multiple indexes attain the maximum value, we assume that $\arg\max{}$ returns the smallest one.
    
    In the rest of this subsection, we prove \cref{prop:minusoptimum} by showing that the optimal solution is located at a vertex of the polyhedron determined by the inequality constraints.
    
    \begin{proof}[Proof of \cref{prop:minusoptimum}]
        First, we show that a global optimum is represented by \cref{eq:vartex} for some $\optconstminus\in\setR[]$.
        Let $\yminus[]=\paren*{\yminus[1],\dots,\yminus[\numminus]}$ and let $\objfuncminus\paren*{\yminus[]}$ be the objective function. Suppose that
        \begin{align}
        \label{eq:nonvertex}
            &\yminus[1]\geq \dots  \geq \yminus[L_1-1]
            > \yminus[L_1] =\dots =  \yminus[L_2]
            > \yminus[L_2+1] = \dots = \yminus[\numminus] = 0
        \end{align}
        for integers $L_1$ and $L_2$ such that $1\leq L_1\leq L_2\leq \numminus$. It is clear that this $\yminus[]$ satisfies the inequality constraints. 
        Let us consider to parameterize $\yminus[]$ as 
        \begin{align*}
            \yminus[]\paren*{t} =
            \begin{cases}
                \yminus -at & \paren*{1\leq \idxminus\leq L_1-1},\\
                \yminus + t & \paren*{L_1\leq \idxminus\leq L_2},\\
                0 & \paren*{L_2+1\leq \idxminus \leq \numminus},
            \end{cases}
        \end{align*}
        where $t$ is a variable that moves across an interval in which the inequality constraints hold and $a>0$ is a constant determined by the equality constraint. 
        We note that $t=0$ is the interior point of the interval.
        Then, we have
        \begin{align*}
            \objfuncminus\paren*{\yminus[]\paren*{t}} = -\sum_{\idxminus=1}^{L_1-1}\paren*{\yminus-at}^\bbeta-\sum_{\idxminus=L_1}^{L_2}\paren*{\yminus+t}^\bbeta.
        \end{align*}
        This function is concave with respect to $t$, because we have
        \begin{align*}
            \frac{\mathrm{d}^2}{\mathrm{d}t^2}\objfuncminus\paren*{\yminus[]\paren*{t}} = -a^2\bbeta\paren*{\bbeta-1}\sum_{\idxminus=1}^{L_1-1}\paren*{\yminus-at}^{\bbeta-2}-\bbeta\paren*{\bbeta-1}\sum_{\idxminus=L_1}^{L_2}\paren*{\yminus+t}^{\bbeta-2}<0,
        \end{align*}
        where we use $\bbeta>1$ for the last inequality.
        This implies that $\objfuncminus\paren*{\yminus[]\paren*{t}}$ takes its minimum value at one endpoint of the interval. Using this argument, we can prove that a global optimum is represented by \cref{eq:vartex} by contradiction. If the optimal value is obtained by $\yminus[]$ that does not satisfy \cref{eq:vartex}, there exist $L_1$ and $L_2$ for which $\yminus[]$ satisfies \cref{eq:nonvertex}. Then, we can continuously transform $\yminus$ while decreasing the function value. This contradicts that $\yminus[]$ is a global optimum.
        
        It remains to specify $\optconstminus$. For $\lminus\in \setIntvl[\numminus]$, the equality constraint implies
        $\paren*{\sum_{\idxminus=1}^{\lminus}\hminus}\optconstminus=\eqval.$
        Hence, it holds that
        $\optconstminus= \eqval / \sum_{\idxminus=1}^{\lminus}\hminus$,
        which gives the conclusion.
    \end{proof}
    
    \subsection{Analysis of middle-level problem \texorpdfstring{\cref{prob:middlevsubprob}}{\space}} \label{subsec:middlelevelanalysis}
    Define $\optvalplus[\numplus]\coloneqq \optvalplus[{\numplus,\transidplus[\numplus]}]$ and $\optvalminus[\numminus]\coloneqq \optvalminus[{\numminus,\lminus[\numminus]}]$ where
    \begin{align}
        &\optvalplus[{\numplus,\transidplus}] \coloneqq \transidplus\cdot
        \sbra*{\paren*{\sum_{\idxplus=1}^{\transidplus} \hplus}^\frac{1}{1-\alpha} \!\!+\!\!\sum_{\idxplus=\transidplus[]+1}^\numplus\paren*{\transidplus^{\alpha}\hplus}^\frac{1}{1-\alpha}}^{-\frac{1-\alpha}{\alpha}}
        ,\label{def:optvalplus}\\
        &\optvalminus[\numminus,\lminus] \coloneqq \lminus \cdot \paren*{\propminus\sum_{\idxminus=1}^{\lminus}\hminus}^{-\frac{1}{\beta}}.
        \label{def:optvalminus}
    \end{align}
    We provide the optimal solution of \cref{prob:middlevsubprob}.
    It follows from \Cref{corolally:posioptval,theo:minus} that \cref{prob:middlevsubprob} reduces to
    \begin{align}
        \min_{\substack{\eqval \in \setRp[]}}{
        \optvalplus[\numplus]\cdot \eqval^{\frac{1}{\alpha}}-\optvalminus[\numminus] \cdot \eqval^{\frac{1}{\beta}}.}
        {\label{prob:reducedmiddle}}
    \end{align}
    
    This is a $1$-dimensional and possibly nonconvex optimization problem.
    The following theorem provides the optimal solution of \cref{prob:reducedmiddle}.
    \begin{theorem}\label{theo:conv}
        The optimal value of \cref{prob:reducedmiddle} is
        \textrm{(i)} $\midoptval$ if $\alpha<\beta$,
        \textrm{(ii)} $0$ if $\alpha = \beta$ and $\optvalplus[\numplus] \geq \optvalminus[\numminus]$, and 
        \textrm{(iii)} $-\infty$ otherwise;
        where 
        \begin{align*}
            \midoptval = -\frac{\paren*{\optvalminus[\numminus]}^\frac{\beta}{\beta-\alpha}}{\paren*{\optvalplus[\numplus]}^\frac{\alpha}{\beta-\alpha}} \paren*{\frac{\alpha}{\beta}}^{\frac{\alpha}{\beta-\alpha}}\paren*{1-\frac{\alpha}{\beta}}.
        \end{align*}
    \end{theorem}
    \begin{proof}
        As for the case $\alpha < \beta$, we can easily check that the global optimum is $\paren*{\frac{\alpha \optvalminus[\numminus]}{\beta \optvalplus[\numplus]}}^\frac{\alpha\beta}{\beta-\alpha}$ by the first derivative test.
        For the other cases, the proofs are straightforward.
    \end{proof}
    
    \cref{theo:conv} demonstrates that the optimal value of \cref{prob:reducedmiddle} strongly depends on the relation of $\alpha$ and $\beta$. 
    If $\alpha < \beta$, indicating that the buyers are more sensitive to minor losses than minor gains, 
    then the optimal value is a non-zero finite one. Otherwise, the seller can make an infinite profit or fail to gain any benefit, which seems inconsistent with reality.
    While the original CPT framework~\citep{tversky1992advances} assumes $\alpha=\beta$, subsequent surveys~\citep{rieger2011prospect,Riegeretal2017estimateCPTparams} have found that the relationship $\alpha < \beta$ holds in many countries.
    
    \begin{remark}
        Similar to \cref{theo:conv}, \cite{azevedo2012risk} also reported that $\alpha=\beta$ may result in an unrealistic situation, where the seller can either make an infinite profit or no profit at all.
        In their analysis, they consider a two-lottery design with the weighting function of the specific form in \cref{eq:weightfunc} and control the probabilities of the winning and losing tickets.
        They further demonstrate that in the case $\alpha>\weightfuncparamplus$ or $\alpha<\weightfuncparamminus$, the seller can achieve an infinite profit.
        Note that, although our result shares similarities, we consider a different setting in which any number of lotteries and any weighting functions are permitted, with the winning probability of each lottery being uniform.
        
    \end{remark}
    
    
    \subsection{Analysis of high-level problem \texorpdfstring{\cref{prob:orgNminusplus}}{\space} and linear-time algorithm}\label{subsec:highlevelanalysis}
    In this subsection, we propose a linear-time algorithm to solve the entire problem of the optimal lottery by analyzing an algorithmic framework for the high-level problem~\cref{prob:orgNminusplus}.
    We assume that basic arithmetic operations (i.e., addition, subtraction, multiplication, division, and exponentiation) can be performed in constant time.
    As we will see in the following, a naive brute-force algorithm requires a quadratic time.
    However, by utilizing the monotonicity of the auxiliary variable, the computation time can be reduced to linear.
    We first give a naive quadratic time algorithm and then improve it to run in linear time. 
\subsubsection{\texorpdfstring{Naive $O\paren*{\N^2}$-time algorithm}{}}
    This naive algorithm computes the optimal solution of \cref{prob:orgNminusplus} by the brute-force search for $\paren*{\numminus,\numplus}\in\setN[2]$ under $\numminus+\numplus=\N$.
    Under each fixed $\paren*{\numminus,\numplus}$, it computes the optimal value of \cref{prob:middlevsubprob} as follows:
    it first computes $\transidplus[\numplus]$ and $\lminus[\numminus]$ by \cref{eq:transiddef,def:lminusnumminus}, respectively.
    Then, it computes $\optvalplus[\numplus]$ and $\optvalminus[\numminus]$ by \cref{def:optvalplus,def:optvalminus} and the optimal value under the fixed $\paren*{\numminus,\numplus}$ as \cref{theo:conv}.
    Since the computation of $\optvalplus[\numplus]$ and $\optvalminus[\numminus]$ 
    takes $O\paren*{\N}$ time and the algorithm computes them at every $\paren*{\numminus, \numplus}$, the overall time complexity is $O\paren*{\N^2}$.

\subsubsection{\texorpdfstring{Improved $O\paren*{\N}$-time algorithm}{Improved O(N)-time algorithm}}
    The bottleneck of the above algorithm is the calculation of $\transidplus[\numplus]$, $\lminus[\numminus]$, $\optvalplus[\numplus]$, and $\optvalminus[\numminus]$.
    First, we provide an efficient subroutine to compute $\paren*{\transidplus[1], \optvalplus[1]}, \ldots, \paren*{\transidplus[\N], \optvalplus[\N]}$ all at once in $O\paren*{\N}$ time.
    In this subsection, we use the symbol $\hplus[\idxplus,\numplus]$ to denote $\hplus$ when there are $\numplus$ gain tickets, i.e., 
    \begin{align*}
        \hplus[\idxplus,\numplus]= \weightfuncplus\paren*{\frac{\numplus-\idxplus+1}{\N}} - \weightfuncplus\paren*{\frac{\numplus-\idxplus}{\N}}.
    \end{align*}
    Note that, by definition, it holds that 
    \begin{align}\label{eq:hplusNpNpone}
        \hplus[\idxplus,\numplus]=\hplus[\idxplus-1,\numplus-1]
    \end{align}
    for any $\numplus\in \sbra*{\N}\setminus\brc*{1}$ and $\idxplus\in\sbra*{\numplus}\setminus\brc*{1}$.
    We first derive the monotonicity of $\paren*{\transidplus[\idxparameterized]}_{\idxparameterized\in\sbra*{\N}}$.
    \begin{lemma}
        \label{lemm:Jnondecrease}
        $\paren*{\transidplus[\idxparameterized]}_{\idxparameterized\in\sbra*{\N}}$ is monotonically nondecreasing.
    \end{lemma}
    \begin{proof}
        We only need to show $\transidplus[\idxparameterized+1]\geq\transidplus[\idxparameterized]$ when $\transidplus[\idxparameterized]\geq 2$ holds.
        It follows from the definition \cref{eq:transiddef} that
        \begin{align}
            \label{eq:Jnondecrease1}
            \paren*{\transidplus[\idxparameterized]-1} \cdot \hplus[{\transidplus[\idxparameterized]},\idxparameterized] & < \sum_{\idxplus=1}^{{\transidplus[\idxparameterized]}-1}\hplus[\idxplus,\idxparameterized].
        \end{align}
        The inequality implies that 
            $\hplus[{\transidplus[\idxparameterized]},\idxparameterized] < \hplus[1,\idxparameterized]$;
        indeed, if $\hplus[{\transidplus[\idxparameterized]},\idxparameterized]\geq \hplus[1,\idxparameterized]$ holds, then, it follows from \cref{defi:peakindex} and $\transidplus[\idxparameterized] \geq \flexplus$ that $\hplus[{\transidplus[\idxparameterized]},\idxparameterized] \geq \hplus[\idxplus,\idxparameterized]$ for all $\idxplus\in\sbra*{\transidplus[\idxparameterized]-1}$, which contradicts to \cref{eq:Jnondecrease1}.
        
        Based on the result, we have
        \begin{align}
            \transidplus[\idxparameterized] \cdot \hplus[{\transidplus[\idxparameterized]}+1,\idxparameterized+1] = \transidplus[\idxparameterized] \cdot \hplus[{\transidplus[\idxparameterized]},\idxparameterized] &< \hplus[{\transidplus[\idxparameterized]}+1,\idxparameterized+1] + \sum_{\idxplus=1}^{{\transidplus[\idxparameterized] - 1}}\hplus[\idxplus,\idxparameterized]
            < \hplus[1,\idxparameterized+1] + \sum_{\idxplus=1}^{{\transidplus[\idxparameterized] - 1}}\hplus[\idxplus+1,\idxparameterized+1] = \sum_{\idxplus=1}^{{\transidplus[\idxparameterized]}}\hplus[\idxplus,\idxparameterized+1],\label{eq:loweraverage}
        \end{align}
        where the first equality follows from \cref{eq:hplusNpNpone}, the first inequality from \cref{eq:Jnondecrease1}, and the second one from \cref{eq:hplusNpNpone} and \cref{eq:Jnondecrease1}.
        
        The inequality \cref{eq:loweraverage} implies that $\transidplus[\idxparameterized+1]\geq \transidplus[\idxparameterized]$. To see this, suppose, for the sake of contradiction, that $\transidplus[\idxparameterized+1]<\transidplus[\idxparameterized]$ holds. 
        By \cref{defi:transididx,defi:peakindex}, we have 
        \begin{align*}
            \transidplus[\idxparameterized+1]\ge\flexplus \text{ and } \frac{\sum_{\idxplus=1}^{\flexplus}\hplus[\idxplus,\idxparameterized+1]}{\flexplus}\le \hplus[\flexplus,\idxparameterized+1]\le \hplus[\flexplus+1,\idxparameterized+1]\le\cdots\le \hplus[\idxparameterized+1,\idxparameterized+1].
        \end{align*}
        Since $\transidplus[\idxparameterized+1] < \transidplus[\idxparameterized] \leq \idxparameterized = \paren*{\idxparameterized+1}-1$ holds, we obtain
       \begin{align*}
           \sum_{\idxplus=1}^{\transidplus[\idxparameterized]}\hplus[\idxplus, \idxparameterized+1] = \flexplus\cdot\frac{\sum_{\idxplus=1}^{\flexplus}\hplus[\idxplus,\idxparameterized+1]}{\flexplus}+\hplus[\flexplus+1,\idxparameterized+1]+\cdots+\hplus[{\transidplus[\idxparameterized]},\idxparameterized+1]
       \le \transidplus[\idxparameterized] \cdot \hplus[{\transidplus[\idxparameterized]}+1,\idxparameterized+1],
       \end{align*}
        which contradicts to \cref{eq:loweraverage}.
    \end{proof}  
    Using this property, we improve the computation of $\paren*{\transidplus[1], \optvalplus[1]}, \ldots, \paren*{\transidplus[\N], \optvalplus[\N]}$ as follows: 
    we compute prefix sums $\genhplus[0],\ldots,\genhplus[\N]$ such that
    \begin{align}
        \genhplus \coloneqq \sum_{\idxprefixary=1}^{\idxgenhplus}\paren*{\weightfuncplus\paren*{\frac{\idxprefixary}{\N}} - \weightfuncplus\paren*{\frac{\idxprefixary-1}{\N}}}^{\frac{1}{1-\alpha}}\label{def:prefixsum}
    \end{align}
    in $O\paren*{\N}$ time. 
    Note that, for each $\idxparameterized\in\setIntvl[\N]$ and $\transidplus\in\setIntvl[\idxparameterized]$, it follows from the definition of $\hplus$ that
    \begin{align}\label{eq:sumhreduced}
        \sum_{\idxplus=1}^{\transidplus[]}\hplus[\idxplus,\idxparameterized] = \weightfuncplus\paren*{\frac{\idxparameterized}{\N}} - \weightfuncplus\paren*{\frac{\idxparameterized-\transidplus[]}{\N}}.
    \end{align}
    Next, we compute $\paren*{\transidplus[\idxparameterized]}_{\idxparameterized\in\sbra*{\N}}$: for each $\idxparameterized$-th iteration, we set $\transidplus[]=\transidplus[\idxparameterized-1]$ and increase $\transidplus$ until it reaches $\idxparameterized$ or it satisfies
    \begin{align}
        \label{eq:algoJtransidcond}
         \weightfuncplus\paren*{\tfrac{\idxparameterized}{\N}}& - \weightfuncplus\paren*{\tfrac{\idxparameterized-\transidplus[]}{\N}}
        \leq\transidplus[] \paren*{\weightfuncplus\paren*{\tfrac{\idxparameterized-\transidplus}{\N}} - \weightfuncplus\paren*{\tfrac{\idxparameterized-\transidplus-1}{\N}}}.
    \end{align}
    From \cref{lemm:Jnondecrease}, the computation takes $O\paren*{\N}$ time.
    We compute each $\optvalplus[\idxparameterized]$ in constant time by using \cref{def:prefixsum,eq:sumhreduced}.
    We formalize the procedure above as \cref{algo:precomputegain}. 
    
    \begin{algorithm}[t]
        \SetAlgoLined
        \DontPrintSemicolon
        \KwIn{Number of lottery tickets $\N$, CPT parameters $\alpha$ and $\weightfuncplus$}
        \KwOut{$\paren*{\transidplus[1], \optvalplus[1]}, \ldots, \paren*{\transidplus[\N], \optvalplus[\N]}$}
        Set $\genhplus[0]\leftarrow 0$;\;
        \lFor{$t\leftarrow 1$ \KwTo $\N$}{
          Set $\genhplus\leftarrow \genhplus[t-1]+\paren*{\weightfuncplus\paren*{\frac{t}{\N}} - \weightfuncplus\paren*{\frac{t-1}{\N}}}^{\frac{1}{1-\alpha}}$;
        }
        Set $\transidplus[]\leftarrow 1$;\;
        \For{$\idxparameterized \leftarrow 1$ \KwTo $\N$}{
            \lWhile{$\transidplus<\idxparameterized$ and \cref{eq:algoJtransidcond} fails}{$\transidplus[] \leftarrow \transidplus[]+1$;}
            Store $\transidplus[\idxparameterized]\leftarrow\transidplus[]$ and
            $\optvalplus[\idxparameterized] \leftarrow \tfrac{\transidplus[]}{\sbra*{ \paren*{W\paren*{\frac{\idxparameterized}{\N}}-W\paren*{\frac{\idxparameterized-\transidplus[]}{\N}}}^{\frac{1}{1-\alpha}}+\transidplus^{\frac{\alpha}{1-\alpha}}\genhplus[k-J]}^{\frac{1-\alpha}{\alpha}}}$;\;
            }
        \caption{Precomputing $\paren*{\transidplus[1], \optvalplus[1]}, \dots, \paren*{\transidplus[\N], \optvalplus[\N]}$}
        \label{algo:precomputegain}
    \end{algorithm}
    
    Next, we analyze $\lminus[\numminus]$ and $\optvalminus[\numminus]$ at the optimal solution.
    We can prove that there are no zero components in \\ $\paren*{\wminus[1]^{\ast},\dots,\wminus[\numminus]^{\ast},\allowbreak\wplus[1]^{\ast},\dots,\wplus[\numplus]^{\ast}}$ unless the optimal value of \cref{prob:orgNminusplus} is zero as follows:  
    \begin{proposition}
    \label{prop:optnonzero}
        Let $\paren*{\numminus^{\ast},\numplus^{\ast}}$ be a global optimum solution of \cref{prob:orgNminusplus} and suppose that the optimal value $\combval^{\ast}$ is positive.
        Then, for every $\paren*{\wminus[1]^{\ast},\dots,\wminus[\numminus^{\ast}]^{\ast},\wplus[1]^{\ast},\dots,\wplus[\numplus^{\ast}]^{\ast}}$ that is a global optimum of the problem
        \cref{prob:orgmaxprofit} 
        with $\paren*{\numminus^{\ast},\numplus^{\ast}}$, we have $\wminus[\idxminus]^{\ast}\neq 0$ for any $\idxminus\in\setIntvl[\numminus^{\ast}]$ and $\wplus[\idxplus]^{\ast}\neq 0$ for any $\idxplus\in\setIntvl[\numplus^{\ast}]$.
    \end{proposition} 
    \begin{proof}
        Let $\yoptminus=\U\paren*{\wminus[\idxminus]^{\ast}}$ for $\idxminus\in\sbra*{\numminus^{\ast}}$ and $\yoptplus=\U\paren*{\wplus[\idxplus]^{\ast}}$ for $\idxplus\in\sbra*{\numplus^{\ast}}$. 
        Then, $\paren*{\yoptminus}_{\idxminus\in\sbra*{\numminus^{\ast}}}$ and $\paren*{\yoptminus}_{\idxminus\in\sbra*{\numminus^{\ast}}}$ are respectively global optimum of the problems on losses 
        \cref{cstr:4thlosssubprob}
        and that on gains
        \cref{cstr:4thgainsubprob}
        with $\eqval=\eqval^{\ast}$ where
        \begin{align*}
            \eqval^{\ast}\coloneqq-\sum_{\idxminus=1}^{\numminus}\hminus\U\paren*{\wminus[\idxminus]^{\ast}}=\sum_{\idxplus=1}^{\numplus}\hplus\U\paren*{\wplus[\idxplus^{\ast}]}.
        \end{align*}
        We prove the proposition by contradiction. Suppose that there exists a zero component in $\paren*{\wminus[1]^{\ast},\dots,\wminus[\numminus]^{\ast},\wplus[1]^{\ast},\dots,\wplus[\numplus]^{\ast}}$ (i.e., also in $\paren*{\yoptminus[1],\dots,\yoptminus[\numminus],\yoptplus[1],\dots,\yoptplus[\numplus]}$). Since $\combval^{\ast}>0$, we have $\numminus^{\ast}>0$, $\numplus^{\ast}>0$, and $\eqval^{\ast}>0$.
        By \cref{THEO:PLUS}, we have $\yoptplus>0$ for any $\idxplus\in\setIntvl[\numplus]$, and hence $\wplus[\idxplus]^{\ast}\neq 0$ for any $\idxplus\in\setIntvl[\numplus]$.        
        Therefore, we may assume that $\wminus[\idxminus]^{\ast}=0$ for some $\idxminus\in\sbra*{\numminus}$, which implies $\yoptminus[\numminus]=0$ by \cref{cstr:yminusMonotone}. 
    
        In what follows, we show
        \begin{align}
            \label{eq:nonzero}
            \combval\paren*{\numminus^{\ast}-1,\numplus^{\ast}+1}>\combval\paren*{\numminus^{\ast},\numplus^{\ast}},
        \end{align}
        which contradicts the optimality of $\paren*{\numminus^*,\numplus^*}$.
        To see this inequality, let us consider $\paren*{\yoptminus[1],\dots,\yoptminus[\numminus-1]}$ and $\paren*{\yoptminus[\numminus],\yoptplus[1],\dots,\yoptplus[\numplus]}$, which are respectively feasible solutions of the problem on loss 
        \cref{cstr:4thlosssubprob}
        and that on gains
        \cref{cstr:4thgainsubprob}
        with $\numminus=\numminus^*-1,$ $\numplus=\numplus^*+1$, and $\eqval=\eqval^*$ (recall that $\yoptminus[\numminus]=0$).
        By the feasibility of $\paren*{\yoptminus[1],\dots,\yoptminus[\numminus^*-1]}$, 
        we have 
        $\fminus_{\numminus^*}(\eqval^*)\le \fminus_{\numminus^*-1}(\eqval^*).$
        As $\fplus_{\numplus^*}(\eqval^*)\le \fplus_{\numplus^*-1}(\eqval^*)$ is feasible but is not optimum of the problem
        \cref{cstr:4thlosssubprob}
        by \cref{THEO:PLUS}, we have
        $\fplus_{\numplus^*}(\eqval^*)< \fplus_{\numplus^*-1}(\eqval^*).$
        Combining these inequalities, as for the optimization problem \cref{prob:middlevsubprob} with $\numminus=\numminus^{\ast}-1$ and $\numplus=\numplus^{\ast}+1$, it holds that $\eqval=\eqval^{\ast}$ achieves the objective value smaller than $-\combval\paren*{\numminus^{\ast},\numplus^{\ast}}$, which is the optimal value of \cref{prob:middlevsubprob} with $\numminus=\numminus^{\ast}$ and $\numplus=\numplus^{\ast}$. This concludes \cref{eq:nonzero}. 
    \end{proof}

    Therefore, to obtain the overall optimal solution, we can skip the computation of $\lminus[\numminus]$ by replacing $\lminus[\numminus]$ with $\numminus$ (i.e., replacing $\optvalminus[\numminus]$ with $\optvalminus[\numminus,\numminus]$). 
    With this replacement, we can compute $\optvalminus[\numminus,\numminus]$ in $O\paren*{1}$ time for each $\numminus\in\setIntvl[\N]$ by 
    \begin{align}
    \label{algo:computeoptvalminus}
    \optvalminus[\numminus,\numminus] = \numminus \cdot \paren*{\propminus\sum_{\idxminus=1}^{\numminus}\hminus}^{-\frac{1}{\beta}}=\numminus \cdot \paren*{\propminus\weightfuncminus\paren*{\frac{\numminus}{\N}}}^{-\frac{1}{\beta}},
    \end{align}
    where we use \cref{def:optvalminus} and 
    \begin{align*}
        \sum_{\idxminus=1}^{\numminus}\hminus = \sum_{\idxminus=1}^{\numminus}\paren*{\weightfuncminus\paren*{\frac{\N-\idxminus+1}{\N}} - \weightfuncminus\paren*{\frac{\N-\idxminus}{\N}}}
        =\weightfuncminus\paren*{\frac{\numminus}{\N}}.
    \end{align*}
    
    With the efficient calculation on $\transidplus[\numplus]$, $\lminus[\numminus]$, $\optvalplus[\numplus]$, and $\optvalminus[\numminus]$, we compute the optimal value of \cref{prob:orgNminusplus} in $O\paren*{\N}$ time: for each $\numplus$, we can calculate $\optval\paren*{\numminus, \numplus}$ in $O\paren*{1}$ time by \cref{theo:conv} with \cref{algo:computeoptvalminus} and precomputed $\optvalplus[\numplus]$.
    Thus, by the linear search with respect to $\numplus$, we can obtain the optimal value of \cref{prob:orgNminusplus} and the corresponding solution in $O\paren*{\N}$ time. 
    We formalize the overall computation for the optimal design of lottery in \cref{algo:optlot}.
    The following proposition formally states the time complexity of the algorithm.
    \begin{proposition}
        \label{prop:complexity}
        \cref{algo:optlot} runs in $O\paren*{\N}$ time.
    \end{proposition}
    \begin{proof}
        Recall that we can compute $\optvalminus[1,1],\dots,\optvalminus[\N,\N]$ in $O\paren*{\N}$ time by \cref{algo:computeoptvalminus}.
        Since \cref{algo:precomputegain} runs in $O\paren*{\N}$ time, we conclude that \cref{algo:optlot} also does in $O\paren*{\N}$ time.
    \end{proof}
    
    Based on \cref{algo:optlot}, we conduct numerical experiments with specific hyperparameters $\paren*{\N, \alpha, \beta, \propminus, \weightfuncplus, \weightfuncminus}$ and give optimal lotteries; see \cref{sec:experiment} for the results.
    
        \begin{algorithm}[t]
            \SetAlgoLined
            \DontPrintSemicolon
            \KwIn{\mbox{Number of lottery tickets $N$, CPT parameters $\alpha$, $\beta$, $\lambda$, $\weightfuncminus$, $\weightfuncplus$}}
            \KwOut{$\combval^{\ast}, \numplus^{\ast},\numminus^{\ast}$, $\wplus[1]^{\ast},\ldots,\wplus[\numplus^{\ast}]^{\ast}$,  $\wminus[1]^{\ast},\ldots,\wminus[\numminus^{\ast}]^{\ast}$}
            Compute $\transidplus[\idxplus]$, $\optvalplus[\idxplus]$ for all $\idxplus\in\sbra*{\N}$ by \cref{algo:precomputegain};\;\label{line:core} 
            Set $\combval^{\ast}\leftarrow 0$ and $\paren*{\numplus^{\ast},\numminus^{\ast}} \leftarrow \paren*{\N,0}$;\;
            \For{$\numplus \leftarrow 1$ \KwTo $\N-1$}{
                Set $\numminus \leftarrow \N-\numplus$ and $\optvalminus[\numminus]\leftarrow \numminus \cdot \paren*{\propminus\weightfuncminus\paren*{\frac{\numminus}{\N}}}^{-\frac{1}{\beta}}$;\; 
                Compute $\combval\paren*{\numminus,\numplus}$ by \cref{theo:conv};\;
                \lIf{
                    $\combval\paren*{\numminus,\numplus}<\combval^{\ast}$}{
                    $\combval^{\ast}\leftarrow \combval\paren*{\numminus,\numplus}$ and $\paren*{\numminus^{\ast},\numplus^{\ast}}\leftarrow \paren*{\numminus,\numplus}$; 
                    }
            }
            Compute $\yoptplus[] \in \setR[\numplus^{\ast}]$ by \cref{eq:thmyoptplus} and $\yoptminus[] \in \setR[\numminus^{\ast}]$ by \cref{eq:vartex};\; 
            Set $\wplus[\idxplus]^{\ast} \leftarrow \paren*{\yoptplus}^{\frac{1}{\alpha}}~\paren*{\forall \idxplus\in\setIntvl[\numplus^{\ast}]}$ and
            $\wminus[\idxminus]^{\ast} \leftarrow - \paren*{\yoptminus/\propminus}^{\frac{1}{\beta}}$ $\paren*{\forall \idxminus\in\setIntvl[\numminus^{\ast}]}$;\;
            \caption{Optimal design of the lottery}
            \label{algo:optlot}
    \end{algorithm}
    
    
    \subsection{Interpretation of the Optimal Lottery Structure}
    \label{sec:interpretation}
    
    We briefly summarize the structural implications of the analysis in \cref{sec:analysis}
    and discuss how the optimal lottery qualitatively depends on the CPT parameters.
    
    \paragraph{Structure of the optimal lottery.}
    Using \cref{THEO:PLUS,prop:minusoptimum}, \cref{prop:optnonzero}, and the relationship $\yminus=-\U\paren*{\wminus[\idxminus]}$ for $\idxminus \in \setIntvl[\numminus]$ and $\yplus=\U\paren*{\wplus[\idxplus]}$ for $\idxplus \in \setIntvl[\numplus]$, the optimal lottery can be summarized as follows.
    There exist $\numminus \ge 1$ and $\transidplus\in\setIntvl[\numplus]$
    such that the optimal outcomes $\paren*{\wconb[\ell]}_{\ell=1}^\N$ satisfy
    \begin{align}\label{eq:summaryoptimallottery}
    \wconb[\ell] =
    \begin{cases}
        -\paren*{\frac{\optconstminus}{\propminus}}^{\frac{1}{\beta}}, & \paren*{1 \le \ell \le \numminus},\\
        Y^{\frac{1}{\alpha}}, & \paren*{\numminus+1 \le \ell \le \numminus+\transidplus}, \\
        \paren*{\tfrac{\transidplus\hplus}{\sum_{\idxplus'=1}^{\transidplus}\hplus[\idxplus']}}^{\frac{1}{1-\alpha}} Y^{\frac{1}{\alpha}}, & \paren*{\numminus+\transidplus+1 \le \ell \le \N},
    \end{cases}
    \end{align}
    where $\bar{Y}$ and $Y$ are positive constants.
    In words, the optimal lottery consists of three distinct parts: outcomes with a constant negative value, gain outcomes with a constant positive value, and a set of gain outcomes with strictly increasing prizes.
    
    \paragraph{Dependence on CPT parameters.}
    The qualitative structure above is shaped by the CPT parameters in intuitive ways.
    The curvature parameters $\alpha$ and $\beta$ directly affect the low-level problems and influence the relative sizes of the flat and monotone regions on the gain and loss sides.
    Stronger curvature tends to enlarge the flat regions and emphasize extreme outcomes.
    The loss aversion parameter $\propminus$, in contrast, does not alter the qualitative shape of the optimal lottery.
    Instead, it enters the middle-level problem and rescales loss outcomes relative to gains, thereby affecting the balance between gains and losses without changing the internal structure on each side.
    
    \section{Lottery design with ticket price constraint}\label{sec:lotterywithpriceconstraint} 
    \label{subsec:optlotpricecstr}
    In the previous sections, the worst outcome can take an arbitrarily large value in some cases.  
    As noted in \cref{sec:formulation}, under our formulation, the worst outcome can be interpreted as a ticket price, in the sense that $-\min_{i \in [N]} u_i$ represents the maximum payment a buyer may incur.
    In this section, we analyze the setting where a cap on the ticket price is externally imposed, aiming to yield more practical results. 
    We first provide its formulation and then propose an algorithm for solving this problem. 
    Moreover, we propose a linear-time algorithm that outputs the optimal solution in a specific case. 
    
    \subsection{Problem formulation and characterization of the optimal solution}
    Let $-\wminus[\min]$ denote the ticket price cap. Then, the constraint $\wminus[1] \geq \wminus[\min]$ is added to the middle-level problem 
    \cref{prob:orgmaxprofit},
    which results in the following low-level problem instead of 
    the problem 
    \cref{cstr:4thlosssubprob}:
    \begin{align}
        \begin{split}\label{prob:minus2}
            \min_{\substack{\yminus[] \in\setR[\numminus]}}\quad&{\sum_{\idxminus=1}^{\numminus}\U^{-1}\paren*{-\yminus}}\\
        \subjectto\quad&{\ymin\ge \yminus[1]\ge \yminus[2]\ge \cdots\ge \yminus[\numminus]\ge 0,}\\	
        &{\sum_{\idxminus=1}^{\numminus}\hminus\yminus=\eqval}
        \end{split}
    \end{align}
    with $ \ymin \coloneqq -\U\paren*{\wminus[\min]}$.
    We can also characterize the optimal value of \cref{prob:minus2} as follows:
    
    \begin{theorem}
        \label{theo:lossbounded}
        The problem \cref{prob:minus2} is feasible if and only if $0 \leq v\leq \ymin\sum_{\idxminus=1}^{\numminus}\hminus$ holds. Under the condition, the optimal value of \cref{prob:minus2} is given by 
        \begin{subequations}\label{eq:optvalminusbounded}
            \begin{align}
            	\min_{\substack{\ell_1,\ell_2\in \setNz}}
            	\quad&{-\ell_1\paren*{\frac{\ymin}{{\propminus}}}^{\frac{1}{\beta}}-\paren*{\ell_2-\ell_1}\paren*{\tfrac{v-\ymin\sum_{\idxminus=1}^{\ell_1}\hminus }{{\propminus}\sum_{\idxminus=\ell_1+1}^{\ell_2}\hminus}}^{\frac{1}{\beta}\label{obj:optvalminusbounded}}}\\
            	\subjectto\quad&\ymin\sum_{\idxminus=1}^{\ell_1}\hminus \leq \eqval \leq \ymin\sum_{\idxminus=1}^{\ell_2}\hminus,\label{cstr:vinterval}\\
                \quad&\ell_1\le \ell_2 \leq \numminus. \label{cstr:orderellonetwo}
            \end{align}
        \end{subequations}
    
    \end{theorem}

    \cref{theo:lossbounded} enables us to reduce \cref{prob:minus2} to an optimization problem with three variables regardless of the dimension of $\yminus[]$.
    Using the result, we transform the middle-level problem into the following form:
    \begin{align}
        \label{prob:comb2}
        \min_{\ell_1,\ell_2\in\setNz,\, \eqval\in\setR_{\geq 0}} \optvalplus[\numplus]\eqval^{1/\alpha}-\optvalminus[\numminus]'\paren*{\ell_1, \ell_2,\eqval}
        \quad \subjectto~\cref{cstr:vinterval} \text{ and }\cref{cstr:orderellonetwo},
    \end{align}
    where
    \begin{align*}
        \optvalminus[\numminus]'\paren*{\ell_1, \ell_2, \eqval}\coloneqq \ell_1\paren*{\frac{\ymin}{\propminus}}^{\frac{1}{\beta}}+\paren*{\ell_2-\ell_1}\paren*{\tfrac{v-\ymin\sum_{\idxminus=1}^{\ell_1}\hminus }{\propminus\sum_{\idxminus=\ell_1+1}^{\ell_2}\hminus}}^{\frac{1}{\beta}}.
    \end{align*}
    \begin{remark}
        As the numerical experiments of \cref{sec:experiment} indicate, 
        the constraint on ticket price 
        prevents lotteries that involve unrealistic payment amounts;
        without this constraint, specific parameter settings result in scenarios where the optimal lottery consists of one losing ticket and other winning tickets with astronomically high values. 
        In such a situation, the constraint makes the optimal solution avoid the unrealistic structure and become more reasonable.
        Note that the constraint also address the similar issue in the setting of \cite{azevedo2012risk} by preventing the unrealistic outcomes.
    \end{remark}
    
    In the rest of the subsection, we prove \cref{theo:lossbounded}.
    To this end, we first introduce the following proposition, a modification of \cref{prop:minusoptimum}.
    \begin{proposition}\label{prop:yoptminuswithfixedpricecstr}
    Suppose that $v\leq\ymin\sum_{\idxminus=1}^{\numminus}\hminus$.
    There exist a positive integer $\ell_1,\ell_2\in\setIntvl[\numminus]\cup\brc*{0}$ and $0\leq \optconstminus\leq \ymin$ such that the optimal solution of \cref{prob:minus2} is obtained by
    \begin{align*}
        \yminus=
        \begin{cases}\ymin & \paren*{1\leq \idxminus\leq \ell_1},\\ \optconstminus& \paren*{\ell_1+1\leq \idxminus\leq \ell_2},\\ 0 &\paren*{\ell_2+1\leq \idxminus\leq \numminus}, \end{cases}
    \end{align*}
    where
    \begin{align*}
        \optconstminus = \frac{v-\ymin\sum_{\idxminus=1}^{\ell_1}\hminus }{\sum_{\idxminus=\ell_1+1}^{\ell_2}\hminus}.
    \end{align*}
    \end{proposition}
    \begin{proof}
    While \cref{prop:minusoptimum} states that the optimal solution consists of two values $\optconstminus$ and $0$, this proposition implies that there are three values: $\optconstminus$, $0$, and $\ymin$. Under the inequality constraint of \cref{prob:minus2}, we have $\sum_{\idxminus=1}^{\numminus}\hminus\yminus\le \ymin\sum_{\idxminus=1}^{\numminus}\hminus$.
    Therefore we need $v\le \ymin\sum_{\idxminus=1}^{\numminus}\hminus$ for the feasibility. We abbreviate the rest of the proof since it is essentially equal to that of \cref{prop:minusoptimum}.
    \end{proof}
    
    Using this result, we prove \cref{theo:lossbounded} as follows.
    
    \begin{proof}[Proof of \cref{theo:lossbounded}]
        Under fixed $\ell_1$ and $\ell_2$, together with specifying $\optconstminus$ by the equality constraint, \cref{prob:minus2} comes down to the 1-dimensional optimization problem with respect to $\eqval$. Then, the brute-force search over the optimal values under all-possible $\ell_1$ and $\ell_2$ gives the global optimum, which is represented by 
        \cref{eq:optvalminusbounded}
        in \cref{theo:lossbounded}.
        The range of $\eqval$ is derived by the following argument: under fixed $\ell_1$ and $\ell_2$, it holds that 
        \begin{align*}
            \ymin\sum_{\idxminus=1}^{\ell_1} \hminus\le \sum_{\idxminus=1}^{\numminus}\hminus\yminus\le\ymin\sum_{\idxminus=1}^{\ell_2}\hminus,
        \end{align*}
        and therefore the equality constraint gives 
        \begin{align*}
            \ymin\sum_{\idxminus=1}^{\ell_1} \hminus\le \eqval\le\ymin\sum_{\idxminus=1}^{\ell_2}\hminus.
        \end{align*}
    \end{proof}

    \subsection{Algorithms}\label{subsec:algowithticketconsts}
    According to the above result, we provide the algorithm for optimal lottery design with a fixed ticket price. 
    We first give a naive algorithm and then 
    improve it to run faster. 
    In the improved algorithm, we solve a simple $1$-dimensional optimization problem for quadratic time. 
    Moreover, we give a linear-time algorithm for the case $\alpha\ge \beta$ that ensures the optimal solution. 
    The results of numerical experiments can be seen in \cref{sec:experiment}. 
    
    \subsubsection{\texorpdfstring{Naive algorithm with $O\paren*{\N^3}$ 1D optimizations}{}}
    To obtain the optimal solution of \cref{prob:comb2}, we solve it with respect to $\eqval$ for every fixed $\ell_1,\ell_2\in\setN[2]$ such that $\ell_1\le\ell_2 \leq \numminus$.
    If $\ell_1=\ell_2$ holds, we have $\yminus[1]=\dots \yminus[\ell_1]=y_{\min}$ and $\yminus[\ell_1+1]=\dots \yminus[\numminus]=0$, and the inequality constraint \cref{cstr:vinterval} reduces to the equality constraint $\eqval = \ymin\sum_{\idxminus=1}^{\ell_1}\hminus$.
    Consequently, the optimal value is achieved at this value of $\eqval$.
    If $\ell_1<\ell_2$, let 
    \begin{align*}
        g_{\ell_1,\ell_2}\paren*{\eqval}&\coloneqq \optvalplus[\numplus]\eqval^{\frac{1}{\alpha}}-\optvalminus[\numminus]'\paren*{\ell_1, \ell_2,\eqval} = \optvalplus[\numplus]\eqval^{\frac{1}{\alpha}}-\paren*{\ell_2-\ell_1}\paren*{\tfrac{v-\ymin\sum_{\idxminus=1}^{\ell_1}\hminus }{{\propminus}\sum_{\idxminus=\ell_1+1}^{\ell_2}\hminus}}^{\frac{1}{\beta}}-\ell_1\paren*{\frac{\ymin}{{\propminus}}}^{\frac{1}{\beta}}.
    \end{align*}
    Letting 
    \begin{align*}
        \kappa\coloneqq \frac{\eqval-\ymin\sum_{\idxminus=1}^{\ell_1}\hminus}{\ymin\sum_{\idxminus=\ell_1+1}^{\ell_2}\hminus},
    \end{align*}
    we have that
    \begin{align*}
        g_{\ell_1,\ell_2}\paren*{\eqval}&=\optvalplus[\numplus]\ymin^\frac{1}{\alpha}\paren*{\sum_{\idxminus=1}^{\ell_1}\hminus+\kappa\sum_{\idxminus=\ell_1+1}^{\ell_2}\hminus}^\frac{1}{\alpha}
        -\paren*{\ell_2-\ell_1}\paren*{\frac{\ymin}{{\propminus}}}^\frac{1}{\beta}\kappa^\frac{1}{\beta}-\ell_1\paren*{\frac{\ymin}{{\propminus}}}^\frac{1}{\beta}.
    \end{align*}
    By ignoring the constant factor $-\ell_1\ymin^\frac{1}{\beta}$ and dividing $g_{\ell_1,\ell_2}\paren*{\eqval}$ by $\optvalplus[\numplus]\ymin^\frac{1}{\alpha}\paren*{\sum_{\idxminus=\ell_1+1}^{\ell_2}\hminus}^{\frac{1}{\alpha}}$, the optimization problem with fixed $\ell_1$ and $\ell_2$ satisfying \cref{cstr:orderellonetwo} reduces to 
    \begin{align}
        \label{prob:comb4}
        \min_{\kappa\in\setR }~\tilde{g}_{\ell_1,\ell_2}\paren*{\kappa}\qquad
        \subjectto~{0\le \kappa\le 1,}
    \end{align} 
    where
    \begin{align*}
        &\tilde{g}_{\ell_1,\ell_2}\paren*{\kappa}
        \coloneqq\paren*{A+\kappa}^{\frac{1}{\alpha}}-B\kappa^{\frac{1}{\beta}},\quad
        A\coloneqq \tfrac{\sum_{\idxminus=1}^{\ell_1}\hminus}{\sum_{\idxminus=\ell_1+1}^{\ell_2}\hminus},\quad
        B\coloneqq \tfrac{\paren*{\ell_2-\ell_1}\ymin^{\frac{1}{\beta}-\frac{1}{\alpha}}}{\optvalplus[\numplus]{\propminus^{\frac{1}{\beta}}}\paren*{\sum_{\idxminus=\ell_1+1}^{\ell_2}\hminus}^{\frac{1}{\alpha}}}.
    \end{align*}
    We note that $A$, $B>0$ by their definition. 
    
    We can prove that this problem has only three candidates as a global optimum: the two endpoints of the constraints (i.e., $\kappa=0,1$) and the local minimum obtained by the first derivative tests. 
    Hence, we can efficiently compute the optimal solution of \cref{prob:comb4} 
    by numerical algorithms 
    such as the Newton method.
    Since there are at most $O\paren*{\numminus^2}$ values of $\paren*{\ell_1,\ell_2}\in\setN[2]$ to consider, we numerically obtain a global optimum of \cref{prob:comb2} by solving $O\paren*{\numminus^2}$ instances of \cref{prob:comb4}. 
    In addition, by considering every possible $\numplus\in\setIntvl[\N]$, we can find the optimal lottery with a fixed price by solving $O\paren*{\N^3}$ instances of \cref{prob:comb4} 
    in total. Note that we can compute the optimal $\yminus[]$ from the optimal solution of \cref{prob:comb2};
    see \cref{prop:yoptminuswithfixedpricecstr} for the detail. 
    
    \subsubsection{\texorpdfstring{Improved algorithm with $O\paren*{\N^2}$ 1D optimizations}{}}
    As with the analysis of the improved algorithm in \cref{subsec:middlelevelanalysis}, we can conclude that there are no zero components in the optimal solution unless the optimal value is zero. 
    Therefore, it suffices to check the case $\ell_2=\numminus$, which reduces the computational cost. 
    Formally, we obtain the following. 
    
    \begin{proposition}\label{prop:fixed-improved}
        The optimal solution of the lottery optimization with the ticket price constraint can be obtained by solving $O\paren*{\N^2}$ instances of \cref{prob:comb4}. 
    \end{proposition}
    \begin{proof}
        Since there are no zero components in the optimal solution unless the optimal value is zero, we only need to consider the case $\ell_2=\numminus$ for each $\paren*{\numplus,\numminus}$.
        Therefore, we only need to solve \cref{prob:comb4} 
        for $\ell_1\in\setIntvl[\numplus]$ and $\ell_2=\numminus$ to obtain the optimal solution, and the total number of the instances to be solved is 
        $\sum_{\numplus=1}^{N}\numplus=\frac{N\paren*{N+1}}{2}=O\paren*{N^2}$.
    \end{proof}
    
\subsubsection[Improved algorithm]{Improved $O\paren*{\N}$-time algorithm for $\alpha \ge \beta$} 
    
    The bottleneck of the above algorithm is the linear search for $\ell_1$ and solving \cref{prob:comb2}. 
    We improve the bottleneck by proving that we only need to consider the case $\yminus=\ymin$ for any $\idxminus\in\setIntvl[\numminus]$ when $\alpha\ge \beta$ holds.
    To this end, we first prove the following lemma.
    \begin{lemma}
        \label{lemm:optyminus}
        Suppose that $\alpha\ge \beta$. 
        Then, for any $\ell_1$, $\ell_2$ satisfying \cref{cstr:orderellonetwo}, the optimal solution of \cref{prob:comb4} satisfies the inequality constraint \cref{cstr:vinterval} with the equality, i.e., the optimal solution $\eqval^*$ satisfies
        \begin{align}
        \label{eq:voptimalequality}
            \eqval^*\in\brc*{\ymin\sum_{\idxminus=1}^{\ell_1}\hminus,\,\ymin\sum_{\idxminus=1}^{\ell_2}\hminus}.
        \end{align}
    \end{lemma}
    \begin{proof}
        When $\ell_1=\ell_2$ holds, the assertion immediately follows, as the inequality constraint \cref{cstr:vinterval} reduces to the equality constraint.
        For the case $\ell_1<\ell_2$,
        we first show that $\tilde{g}'_{\ell_1,\ell_2}\paren*{\kappa}=0$ for $0<\kappa<1$ implies $\tilde{g}''_{\ell_1,\ell_2}\paren*{\kappa}<0$.
        A straightforward calculation shows that
        \begin{align*}
            \tilde{g}'_{\ell_1,\ell_2}\paren*{\kappa} &= \balpha\paren*{A+\kappa}^{\balpha-1}-\bbeta B\kappa^{\bbeta-1},\\
            \tilde{g}''_{\ell_1,\ell_2}\paren*{\kappa} &= \balpha\paren*{\balpha-1}\paren*{A+\kappa}^{\balpha-2}-\bbeta\paren*{\bbeta-1} B\kappa^{\bbeta-2}.
        \end{align*}
        Hence, $\tilde{g}'_{\ell_1,\ell_2}\paren*{\kappa}=0$ (i.e., $\paren*{A+\kappa}^{\balpha-1}=\frac{\bbeta}{\balpha}B\kappa^{\bbeta-1}$) implies that
        \begin{align*}
            \paren*{A+\kappa}\tilde{g}''_{\ell_1,\ell_2}\paren*{\kappa}
            &= \balpha\paren*{\balpha-1}\paren*{A+\kappa}^{\balpha-1}-\bbeta\paren*{\bbeta-1} \kappa^{\bbeta-2}B\paren*{A+\kappa}\\
            &= \balpha\paren*{\balpha-1}\cdot\frac{\bbeta}{\balpha}B\kappa^{\bbeta-1}-\bbeta\paren*{\bbeta-1}B \kappa^{\bbeta-2}\paren*{A+\kappa}\\
            &= \bbeta B\kappa^{\bbeta-2}\big[\paren*{\balpha-\bbeta}\kappa-\paren*{\bbeta-1}A\big]<0,
        \end{align*}
        where the last inequality follows from $\balpha\le\bbeta$ (by $\alpha\ge\beta$) and $\bbeta>1$. 
        Combining this inequality and $A+\kappa> A>0$ gives $\tilde{g}''_{\ell_1,\ell_2}\paren*{\kappa}<0$. 
        
        Now, we prove the lemma by contradiction. 
        Suppose that the optimal value satisfies        
        \begin{align*}
            \eqval^*\notin\brc*{\ymin\sum_{\idxminus=1}^{\ell_1}\hminus,\ \ymin\sum_{\idxminus=1}^{\ell_2}\hminus}.
        \end{align*}
        Then, the optimal solution of \cref{prob:comb4}, denoted by $\kappa^*$, satisfies $0<\kappa^*<1$. 
        Since $\tilde{g}_{\ell_1,\ell_2}$ is continuously differentiable  
        in $\paren*{0,1}$, it must hold that $\tilde{g}'_{\ell_1,\ell_2}\paren*{\kappa^*}=0$. 
        Nevertheless this implies $\tilde{g}''_{\ell_1,\ell_2}\paren*{\kappa^*}<0$, which means $\tilde{g}_{\ell_1,\ell_2}\paren*{\kappa^*}$ is a local maximum. 
        This is the contradiction.
    \end{proof}

    Now, we prove the following proposition.
    \begin{proposition}\label{prop:optyminus}
        Let $\paren*{\numminus^{\ast},\numplus^{\ast}}$ be a global optimum solution of \cref{prob:orgNminusplus} under the ticket price constraint and suppose that the optimal value $\combval^{\ast}$ is positive.
        Then, if $\alpha\ge\beta$ holds, for every $\paren*{\wminus[1]^{\ast},\dots,\wminus[\numminus^{\ast}]^{\ast},\wplus[1]^{\ast},\dots,\wplus[\numplus^{\ast}]^{\ast}}$ that is a global optimum of the problem
        \cref{prob:orgmaxprofit}
        with the additional constraint $\wminus[1]\ge\wminus[\min]$ under $\paren*{\numminus,\numplus}=\paren*{\numminus^{\ast},\numplus^{\ast}}$, we have $\wminus[\idxminus]^{\ast}=\wminus[\min]$ for any $\idxminus\in\setIntvl[\numminus^{\ast}]$.
    \end{proposition}
    \begin{proof}
        By \cref{lemm:optyminus}, for the global optimum of the problem
        \cref{prob:orgmaxprofit},
        the corresponding $\eqval$, $\ell_1$ and $\ell_2$ satisfy the relationship \cref{eq:voptimalequality}. 
        This fact and \cref{prop:yoptminuswithfixedpricecstr} 
        imply that $\yminus^*\in\brc*{\ymin,0}$ for any $\idxminus\in\setIntvl[\numminus^*]$. 
        Moreover, by the same argument as the proof of \cref{prop:optnonzero}, we have $\wminus[\idxminus]^*\neq 0$ ($\Leftrightarrow \yminus^*\neq 0$) for any $\idxminus\in\setIntvl[\numminus^*]$. 
        Thus we obtain that $\yminus^*=\ymin$ for any $\idxminus\in\setIntvl[\numminus^*]$, which gives the conclusion.
    \end{proof}

    We note that $\wminus[\idxminus]^{\ast}=\wminus[\min]$ implies $\yminus^{\ast}=\ymin$. 
    Hence, to obtain optimal $\eqval$ and $\combval\paren*{\numminus,\numplus}$, we can skip the optimization of \eqref{prob:minus2} by just setting $\yminus^{\ast}=\ymin$ for every $\idxminus\in\setIntvl[\numminus]$.
    This fact provides a linear-time algorithm to obtain the optimal solution.
    We formalize the procedure as Algorithm~\ref{algo:optlotpriceconst}. 
    Note that, at the line~\ref{line:core2} of \cref{algo:optlotpriceconst}, we use the equality
    \begin{align*}
        \eqval &= \sum_{\idxminus=1}^\numminus \hminus\ymin
        = \sum_{\idxminus=1}^\numminus\sbra*{\weightfuncminus \paren*{\frac{\idxminus}{\N}} - \weightfuncminus\paren*{\frac{\idxminus-1}{\N}}} \ymin
        =\weightfuncminus\paren*{\frac{\numminus}{\N}}\ymin.
    \end{align*}

    \begin{algorithm}[t]
        \SetAlgoLined
        \DontPrintSemicolon
        \KwIn{Number of tickets $N$, CPT parameters $\alpha$, $\beta$, $\lambda$, $\weightfuncminus$, and $\weightfuncplus$ with $\alpha \ge \beta$, fixed price $-\wminus[\min]$}
        \KwOut{$\combval^{\ast}, \numplus^{\ast},\numminus^{\ast}$,
        $\wplus[1]^{\ast},\ldots,\wplus[\numplus^{\ast}]^{\ast}$, 
        $\wminus[1]^{\ast},\ldots,\wminus[\numminus^{\ast}]^{\ast}$}
        Compute $\transidplus[\idxplus]$, $\optvalplus[\idxplus]$ for all $\idxplus\in\sbra*{\N}$ by \cref{algo:precomputegain};\; 
        Set $\combval^{\ast}\leftarrow 0$ and $\paren*{\numplus^{\ast},\numminus^{\ast}} \leftarrow \paren*{\N,0}$;\;
        \For{$\numplus \leftarrow 1$ \KwTo $\N-1$}{
            Set $\numminus \leftarrow \N-\numplus$ and
            $\eqval\leftarrow \weightfuncminus\paren*{\numminus/\N} \ymin$;\;\label{line:core2}
            $\combval\paren*{\numminus,\numplus} \leftarrow \optvalplus[\numplus] \eqval^{\frac{1}{\alpha}}-\numminus\paren*{\ymin/\lambda}^{\frac{1}{\beta}}$;\;\label{line:core3}
            \lIf{
                $\combval\paren*{\numminus,\numplus}<\combval^{\ast}$}{
         $\combval^{\ast}\leftarrow \combval\paren*{\numminus,\numplus}$ and $\paren*{\numminus^{\ast},\numplus^{\ast}}\leftarrow \paren*{\numminus,\numplus}$; 
                }
        }
        Compute $\yoptplus[] \in \setR[\numplus^{\ast}]$ by \cref{eq:thmyoptplus} and $\yoptminus[] \in \setR[\numminus^{\ast}]$ by \cref{eq:vartex};\; 
        {Set $\wplus[\idxplus]^{\ast} \leftarrow \paren*{\yoptplus}^{\frac{1}{\alpha}}~\paren*{\forall \idxplus\in\setIntvl[\numplus^{\ast}]}$ and $\wminus[\idxminus]^{\ast} \leftarrow - \paren*{\ymin/\lambda}^{\frac{1}{\beta}}$ $\paren*{\forall \idxminus\in\setIntvl[\numminus^{\ast}]}$;}
        \caption{Optimal design of the lottery with the fixed price constraint when $\alpha \geq \beta$}
        \label{algo:optlotpriceconst}
    \end{algorithm}

    We end this subsection by summarizing the computational complexity of the \cref{algo:optlotpriceconst} as follows.
\begin{proposition}
    \label{prop:complexitypriceconst}
    \cref{algo:optlotpriceconst} runs in $O\paren*{\N}$ time.
\end{proposition}
\begin{proof}
    Recall that we can compute $\paren*{\transidplus[1], \optvalplus[1]}, \ldots, \paren*{\transidplus[\N], \optvalplus[\N]}$ and $\optvalminus[1,1],\dots,\optvalminus[\N,\N]$ in $O\paren*{\N}$ time.
    Since for each $\numminus\in\setIntvl[\N]$, the computational time of $\eqval$ and $\combval\paren*{\numminus,\numplus}$ are $O\paren*{1}$, \cref{algo:optlotpriceconst} runs in $O\paren*{\N}$ time.
\end{proof}

\section{Experiments}\label{sec:experiment}
    In this section, we perform numerical experiments to confirm that our algorithms produce optimal lotteries in practical time.
    We do not compare our algorithms with others since no algorithm is known from existing studies that can obtain the optimal lottery due to the differences in the premise of the problem setting.

    We construct concrete optimal lotteries with parameters estimated in \cite{Riegeretal2017estimateCPTparams}, 
    where the value function is \cref{def:utilityfunc} that we use in the theoretical analysis, and the probability weighting functions for gains and losses are \cref{eq:weightfunc}. 
    Several works \citep{rieger2006cumulative,ingersoll2008non,de2012dynamic} have raised the problem that $\weightfuncparamplus,\weightfuncparamminus$ must be larger than $0.279$ to ensure the monotonically increasing probability weighting functions of the form \cref{eq:weightfunc} and all parameters we employ here satisfy this condition.
    All the experiments are implemented in C\texttt{++} and executed on a MacBook Pro 2021 equipped with an Apple M1 Max chip and 64~GB of memory.

    \subsection{Optimal lotteries across countries}

        In this subsection, we present numerical experiments for our proposed methods.
        We first consider the original lottery design problem in \cref{sec:formulation}.
        We then consider the lottery design problem with the ticket price constraint in \cref{sec:lotterywithpriceconstraint}.

    \subsubsection{Original lottery design problem}\label{subsec:lotteryddesignexperiment}

        We first applied our linear-time algorithm (\cref{algo:optlot}) to the original lottery design problem, which consists of the high-level problem~\cref{prob:orgNminusplus}, the middle-level problem~\cref{prob:middlevsubprob}, the low-level problem on losses~\cref{cstr:4thlosssubprob}, and the low-level problem on gains~\cref{cstr:4thgainsubprob}.
        We consider the Canada and the United States settings estimated in \cite{Riegeretal2017estimateCPTparams}.
        To demonstrate scalability, we consider $N=10^{9}$ and show that even very large lotteries can be designed within a practical runtime.

\begin{figure}[t]
          \centering
          \begin{minipage}[t]{0.48\linewidth}
            \vspace{0pt} 
            \centering
            \captionof{table}{Optimal lottery of Canada. 
            Overall odds of gain are 1 in 3.14.}
            \label{tab:canada}
            \centering
            \begin{tabular}{lrl}
                \hline
                Prize & Number & Odds (1 in)\\
                \hline
                $10^8 \sim 10^9$   & $1$         & $1.00 \times 10^9$\\
                $10^7\sim10^8$   & $2$         & $5.00 \times 10^8$\\
                $10^6\sim10^7$   & $34$        & $2.94 \times 10^7$\\
                $10^5\sim10^6$   & $366$       & $2.73 \times 10^6$\\
                $10^4\sim10^5$   & $3871$      & $2.58 \times 10^5$\\
                $10^3\sim10^4$   & $39333$     & $2.54 \times 10^4$\\
                $10^2\sim10^3$   & $360080$    & $2.78 \times 10^3$\\
                $10\sim10^2$     & $2672427$   & $3.74 \times 10^2$\\
                $0\sim10$     & $315227616$ & $3.17$\\
                $-2.30$              & $681696270$ & $1.47$\\
                \hline
            \end{tabular}
          \end{minipage}\hfill
          \begin{minipage}[t]{0.48\linewidth}
            \vspace{0pt} 
            \centering
            \includegraphics[width=\linewidth]{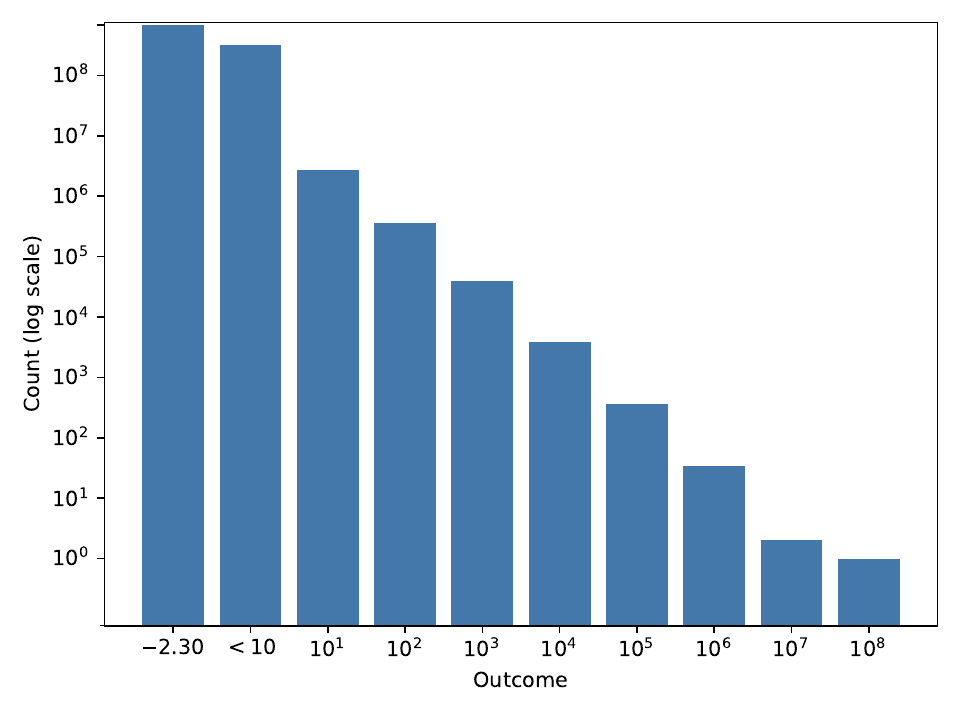}
            \caption{Histogram of optimal lottery for Canada}
            \label{fig:canada}
          \end{minipage}
        \end{figure}

        \paragraph{Canada setting}
        We conducted a numerical experiment with the CPT parameters  $\paren{\alpha,\beta,\propminus,\weightfuncparamplus, \weightfuncparamminus}=(0.42, 0.83, 1.62,\allowbreak 0.44, 0.60)$. 
        The computation time was \textbf{131 seconds}.
        \cref{tab:canada,fig:canada} summarize the optimal design of lotteries.
        The loss, which is the constant negative value in \cref{eq:summaryoptimallottery}, is \$$2.30$, 
        and the maximum gain is \$$1.36 \times 10^{8}$. 
        The ratio of gain tickets to the total number of tickets ($\numplus/\N$) is $31.83\%$,  
        and the seller's profit is \$$7.76 \times 10^{8}$. 
        It shows that in the optimal design, the lottery consists of many small winnings and a few large ones.
        We can say that this result is in line with real lotteries.

\begin{figure}[t]
          \centering
          \begin{minipage}[t]{0.48\linewidth}
            \vspace{0pt}\centering
            \captionof{table}{Optimal lottery of the United States.
            }
            \label{tab:usa}
            \begin{tabular}{lrl}
              \hline
              Prize & Number & Odds (1 in)\\ \hline
              $10^{38}\sim10^{39}$ & $1$ & $1.00 \times 10^9$\\
              $10^{37}\sim10^{38}$ & $10$ & $1.00 \times 10^8$\\
              $10^{36}\sim10^{37}$ & $107$ & $9.35 \times 10^6$\\
              $10^{35}\sim10^{36}$ & $1149$ & $8.70 \times 10^5$\\
              $10^{34}\sim10^{35}$ & $11985$ & $8.34 \times 10^4$\\
              $10^{33}\sim10^{34}$ & $117020$ & $8.55 \times 10^3$\\
              $10^{32}\sim10^{33}$ & $982232$ & $1.02 \times 10^3$\\
              $10^{31}\sim10^{32}$ & $6343123$ & $1.58 \times 10^2$\\
              $10^{30}\sim10^{31}$ & $30652701$ & $32.62$\\
              $10^{29}\sim10^{30}$ & $961891671$ & $1.04$\\
              $-3.53\times 10^{39}$ & $1$ & $1.00 \times 10^9$\\
              \hline
            \end{tabular}
          \end{minipage}\hfill
          \begin{minipage}[t]{0.48\linewidth}
            \vspace{0pt}\centering
            \includegraphics[width=\linewidth]{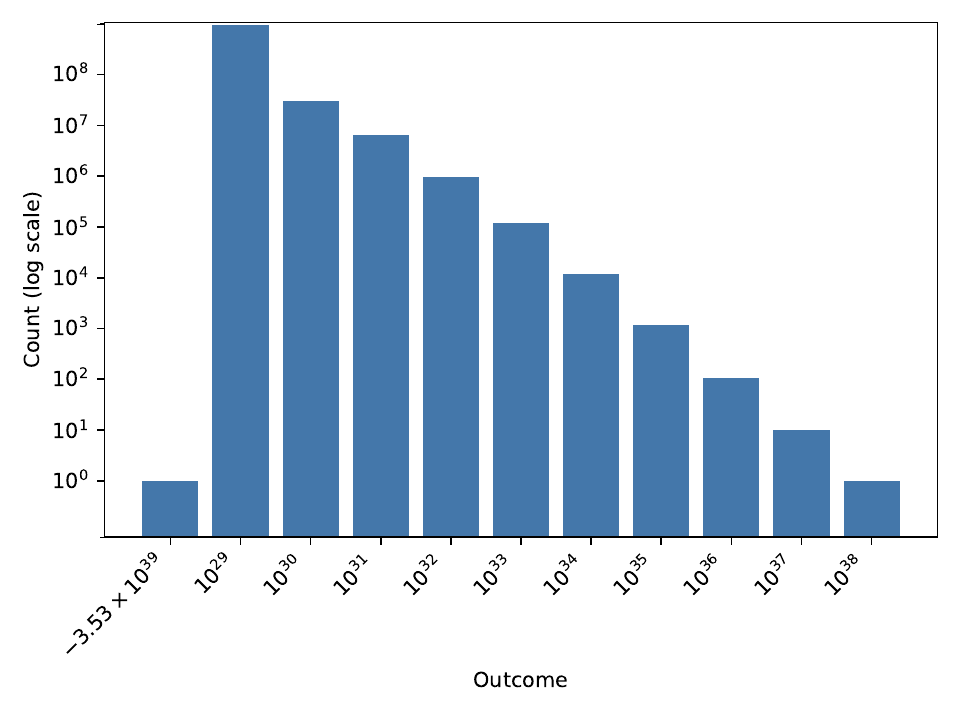}
            \caption{Histogram of optimal lottery for the United States}
            \label{fig:usa}
          \end{minipage}
        \end{figure}

        \paragraph{The United States setting}
        We conducted a numerical experiment with the CPT parameters $\paren{\alpha,\beta,\propminus,\weightfuncparamplus, \weightfuncparamminus}=\paren*{0.42, 0.49, 1.36, 0.44, 0.71}$.
        The computation time was \textbf{123 seconds}.
        \cref{tab:usa,fig:usa} summarize the optimal design of lotteries.
        The loss, which is the constant negative value in \cref{eq:summaryoptimallottery}, is \$$3.53 \times 10^{39}$, 
        and the maximum gain is \$$4.16 \times 10^{38}$. 
        The seller's profit is \$$5.05 \times 10^{38}$. 
        We can see that the optimal lottery consists of tickets that yield gains except for one ticket with an enormous loss. 
        This lottery seems unrealistic because no one has the budget to buy such a too-expensive ticket.
        This result arises from the insensitivity to large losses in the United States setting.
        Since $\beta$ of the United States is $0.49$ and smaller than that of Canada ($\beta=0.83$), we can design the large losses without decreasing the CPT utility much. 
        Similar exorbitant prices can be seen in other settings where $\beta$ is relatively small.
        This phenomenon represents a limitation of the original formulation. Therefore, we explore the lottery design with an additional constraint on the ticket price, as discussed in \cref{subsec:optlotpricecstr}.

    \subsubsection{Lottery design problem with ticket price constraint}
        We consider the optimal design of lottery with the ticket price constraint, in which we use the low-level problem on losses with the ticket price constraint~\cref{prob:minus2} instead of \cref{cstr:4thlosssubprob}.
        We applied our algorithm presented in \cref{prop:fixed-improved} to the United States setting ($\alpha<\beta$) and \cref{algo:optlotpriceconst} to the Greece setting ($\alpha\ge \beta$).
        We set $N=1000$ for the United States setting since the computation requires $O(N^{2})$ time.
        For the Greece setting, we consider $\N=10^{9}$ to demonstrate the scalability of our linear-time algorithm (\cref{algo:optlotpriceconst}) and to show that large lotteries can be designed within a practical runtime.
    
\begin{figure}[t]
          \centering
          \begin{minipage}[t]{0.48\linewidth}
            \vspace{0pt}\centering
            \captionof{table}{Optimal lottery of the United States with the ticket price constraint.
            Overall odds of gain are $1$ in $1.74$.}
            \label{tab:usa-price-cap}
            \begin{tabular}{lrl}
              \hline
              Prize & Number & Odds (1 in)\\ \hline
              $10^2\sim10^3$  & $1$   & $1000$\\
              $10\sim10^2$    & $4$   & $250$\\
              $0\sim10$       & $571$ & $1.75$\\
              $-2$            & $424$ & $2.36$\\
              \hline
            \end{tabular}
          \end{minipage}\hfill
          \begin{minipage}[t]{0.48\linewidth}
            \vspace{0pt}\centering
            \includegraphics[width=\linewidth]{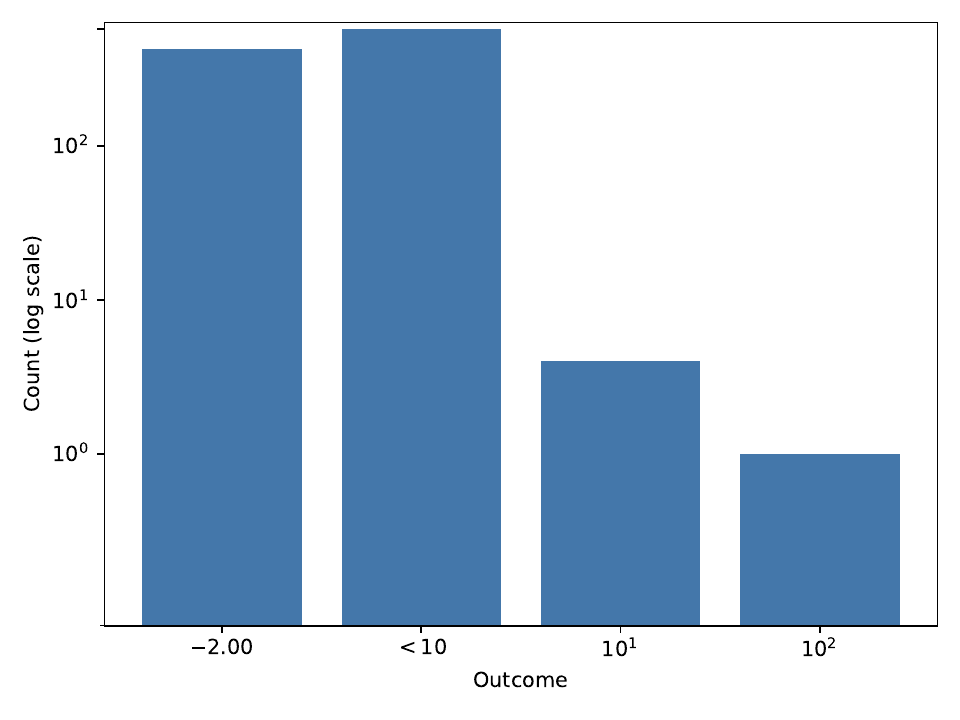}
            \caption{Histogram of optimal lottery for the United States with the ticket price constraint}
            \label{fig:usa-price-cap}
          \end{minipage}
        \end{figure}

        \paragraph{The United States setting}
        We conducted a numerical experiment with $\wminus[\min]=2$, 
        and the same CPT parameters as the case in \cref{subsec:lotteryddesignexperiment}.
        The computation time was \textbf{0.056 seconds}. 
        \cref{tab:usa-price-cap,fig:usa-price-cap} summarize the optimal design of lotteries.
        The worst loss is \$$2$, which coincides with the constraint we imposed, and the maximum gain is \$$310.41$. 
        The ratio of gain tickets to the total number of tickets ($\numplus/\N$) is 57.6\%, 
        and the seller's profit is \$$339.80$. 
        Hence, the additional constraint results in the effective lottery design; the output structure is similar to the Canada setting in \cref{subsec:lotteryddesignexperiment}.

\begin{figure}[t]
          \centering
          \begin{minipage}[t]{0.48\linewidth}
            \vspace{0pt}\centering
            \captionof{table}{Optimal lottery of Greece with the ticket price constraint.
            Overall odds of gain are $1$ in $1.28 \times 10^{3}$.}
            \label{tab:greece-price-cap}
            \begin{tabular}{lrl}
              \hline
              Prize & Number & Odds (1 in)\\ \hline
              $10^7\sim10^8$   & $1$   & $1.00 \times 10^9$\\
              $10^6\sim10^7$   & $5$   & $2.00 \times 10^8$\\
              $10^5\sim10^6$   & $44$  & $2.27 \times 10^7$\\
              $10^4\sim10^5$   & $334$ & $2.99 \times 10^6$\\
              $10^3\sim10^4$   & $2565$  & $3.90 \times 10^5$\\
              $10^2\sim10^3$   & $19139$ & $5.22 \times 10^4$\\
              $10\sim10^2$     & $134410$ & $7.44 \times 10^3$\\
              $0\sim10$        & $624807$  & $1.60 \times 10^3$\\
              $-2$             & $999218695$ & $1.00$\\
              \hline
            \end{tabular}
          \end{minipage}\hfill
          \begin{minipage}[t]{0.48\linewidth}
            \vspace{0pt}\centering
            \includegraphics[width=\linewidth]{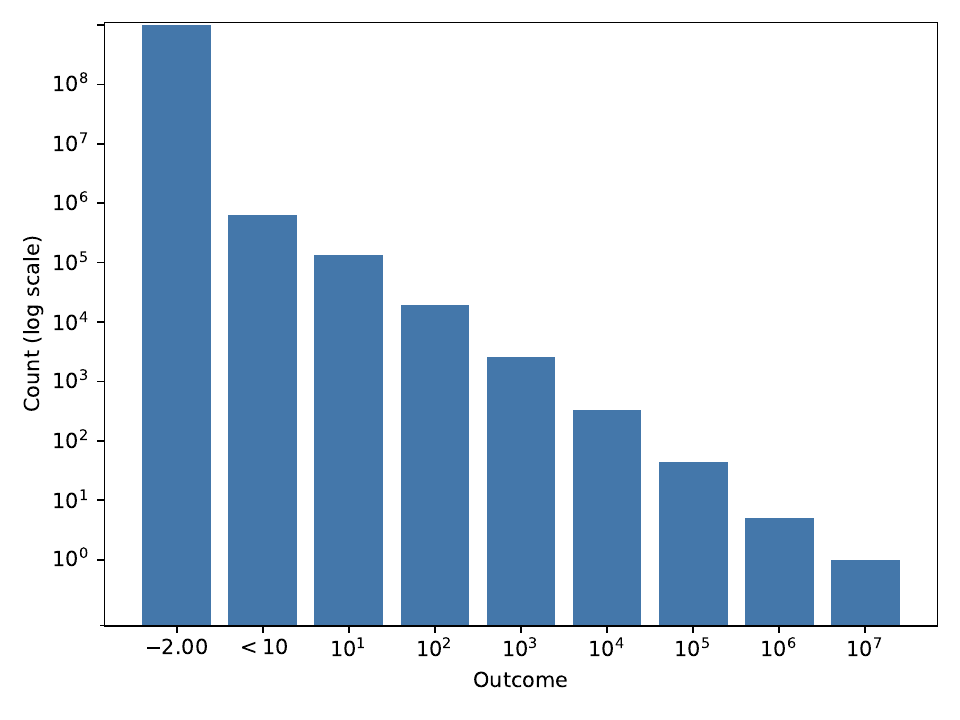}
            \caption{Histogram of optimal lottery for Greece with the ticket price constraint}
            \label{fig:greece-price-cap}
          \end{minipage}
        \end{figure}

        \paragraph{Greece setting} 
        Finally, we conducted a numerical experiment with $\wminus[\min]=2$, and the CPT parameters $\paren{\alpha,\beta,\propminus,\weightfuncparamplus, \weightfuncparamminus}=\paren*{0.50,0.30,1.29,0.44,0.82}$. 
        The computation time was \textbf{96.54 seconds}.
        \cref{tab:greece-price-cap,fig:greece-price-cap} summarize the optimal design of lotteries.
        The loss is \$$2$, which coincides with the constraint we imposed, and
        the maximum gain is \$$4.11\times 10^7$. 
        The ratio of gain tickets to the total number of tickets ($\numplus/\N$) is 0.078\%, 
        and the seller's profit is \$$1.91\times 10^9$. 
        In this setting, the optimal design features a few large gains, while the vast majority of outcomes are losses.

    \subsection{Perturbation analysis for optimal lottery}

        We next examine the robustness of the proposed design: specifically, whether a lottery optimized for a target persona remains attractive enough to be worth purchasing for buyers whose preferences are nearby.
        Let $\alphamean$ and $\betamean$ denote the country-specific estimated means of $\alpha$ and $\beta$, respectively, and $\sigmaalpha$ and $\sigmabeta$ denote the corresponding standard errors, as reported in \cite{Riegeretal2017estimateCPTparams}.
        We first compute an optimal lottery for a given pair of CPT parameters $(\alphamean, \betamean)$ and then assess this lottery in two complementary ways:
        \begin{itemize}
            \item \textbf{Monte Carlo simulation:}
            We independently sample $\alpha \sim \mathcal{N}\paren*{\alphamean, \sigmaalpha^{2}}$ and $\beta \sim \mathcal{N}\paren*{\betamean, \sigmabeta^{2}}$, and compute the buyer's expected utility of the designed lottery for each sampled pair $(\alpha,\beta)$, repeating this procedure $10^{6}$ times.
            \item \textbf{Perturbation heatmap:}
            We compute the buyer's expected utility of the designed lottery for each $(\alpha,\beta) \in \sbra*{\alphamean - \sigmaalpha,\, \alphamean + \sigmaalpha} \times \sbra*{\betamean - \sigmabeta,\, \betamean + \sigmabeta}$ and visualize the resulting values as a heatmap.
        \end{itemize}
        We run these experiments for the four settings considered in the previous section with $\N = 1000$: Canada (unconstrained), the United States (unconstrained), the United States (constrained), and Greece (constrained).

\begin{figure}[t]
          \centering
        
          \begin{subfigure}[t]{\textwidth}
            \centering
            \begin{minipage}[t]{0.49\textwidth}
              \centering
              \includegraphics[width=0.63\linewidth]{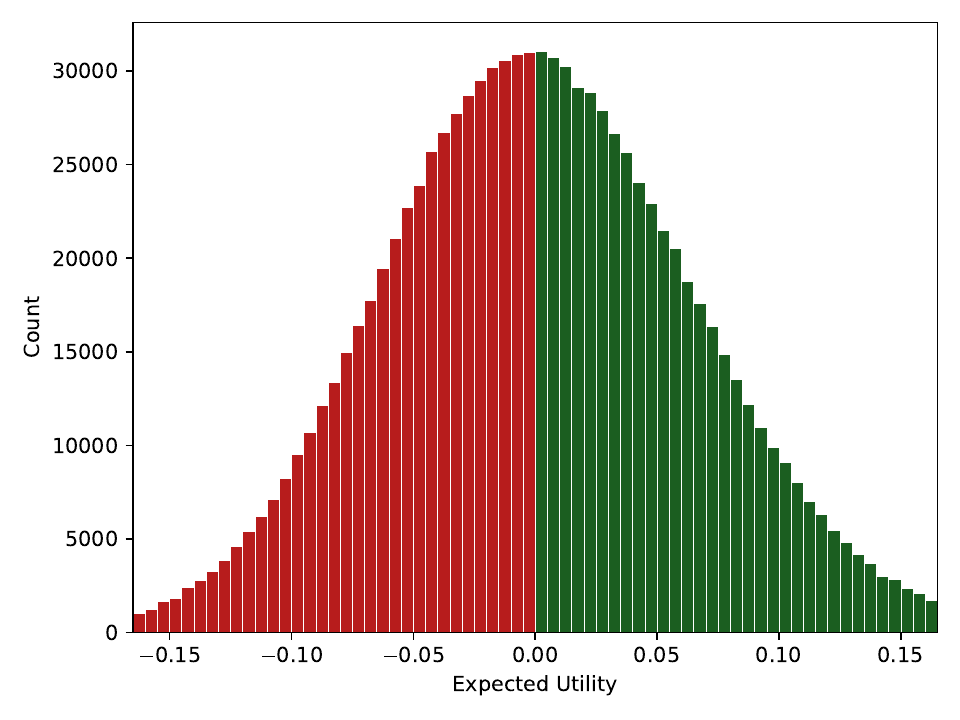}
            \end{minipage}\hfill
            \begin{minipage}[t]{0.49\textwidth}
              \centering
              \includegraphics[width=0.63\linewidth]{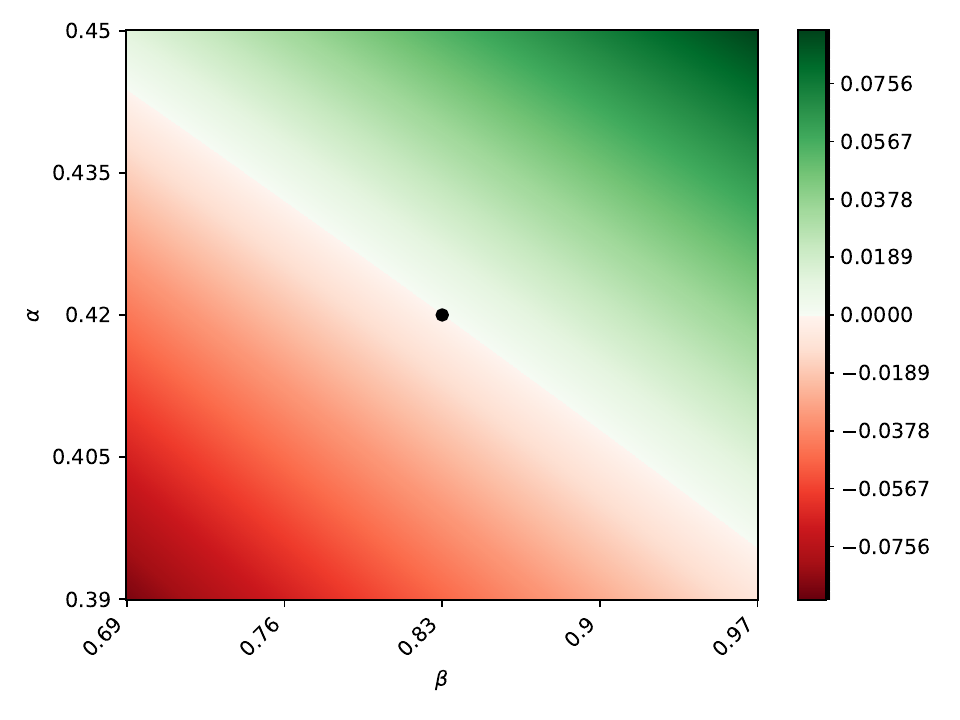}
            \end{minipage}
            \caption{\centering Canada (unconstrained). 
            In the left panel, we exclude 1.31\% of the samples as outliers and use \\ a bin width of $5.00 \times 10^{-3}$.  $\paren{\alphamean\sigmaalpha, \betamean, \sigmabeta, \propminus,\weightfuncparamplus, \weightfuncparamminus}=(0.42, 0.03, 0.83, 0.14, 1.62, 0.44, 0.60)$.}
            \label{fig:country-canada-unc}
          \end{subfigure}
        
          \vspace{3mm}
        
          \begin{subfigure}[t]{\textwidth}
            \centering
            \begin{minipage}[t]{0.49\textwidth}
              \centering
              \includegraphics[width=0.63\linewidth]{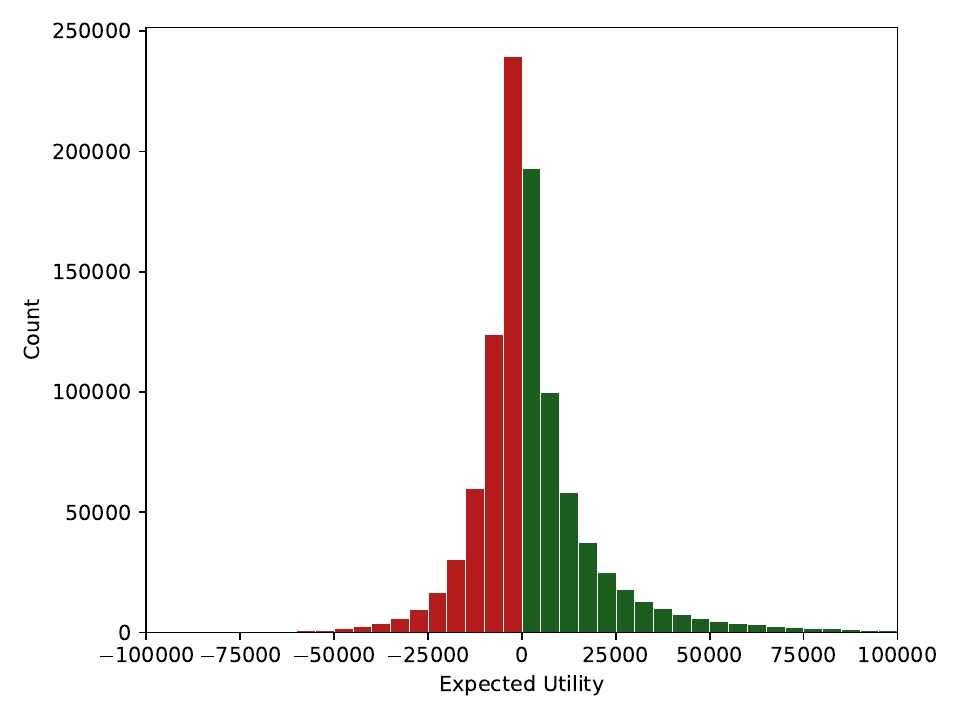}
            \end{minipage}\hfill
            \begin{minipage}[t]{0.49\textwidth}
              \centering
              \includegraphics[width=0.63\linewidth]{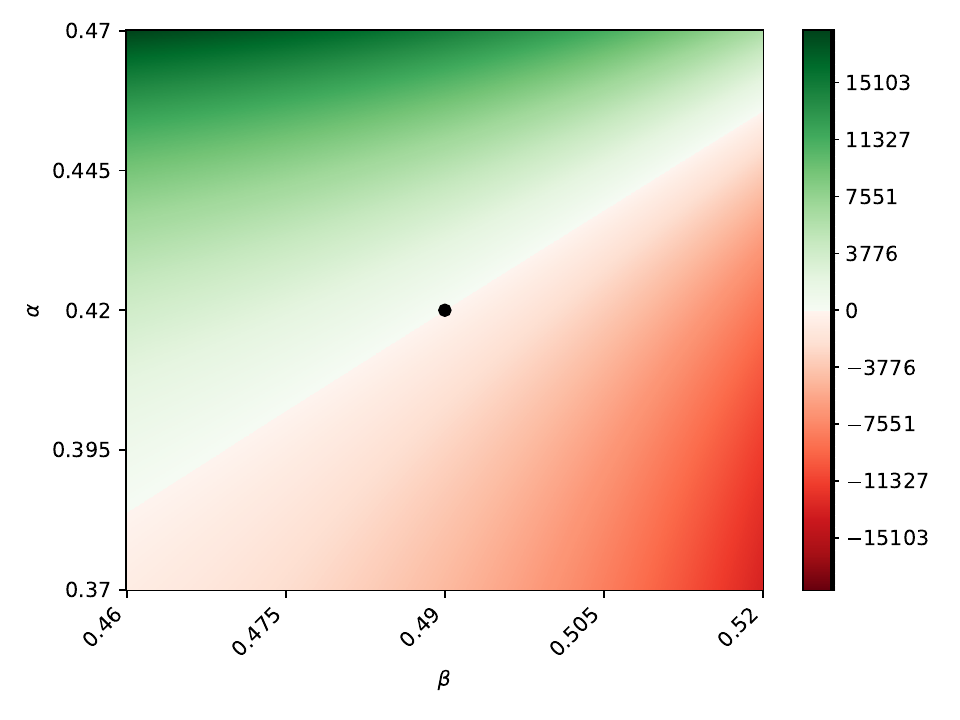}
            \end{minipage}
            \caption{\centering The United States (unconstrained). 
            In the left panel, we exclude 0.94\% of the samples as outliers \\ and use a bin width of $5.00 \times 10^{3}$.
            $\paren{\alphamean, \sigmaalpha, \betamean, \sigmabeta, \propminus,\weightfuncparamplus, \weightfuncparamminus}=\paren*{0.42, 0.05, 0.49, 0.03, 1.36, 0.44, 0.71}$.}
            \label{fig:country-usa-unc}
          \end{subfigure}
        
          \vspace{3mm}
        
          \begin{subfigure}[t]{\textwidth}
            \centering
            \begin{minipage}[t]{0.49\textwidth}
              \centering
              \includegraphics[width=0.63\linewidth]{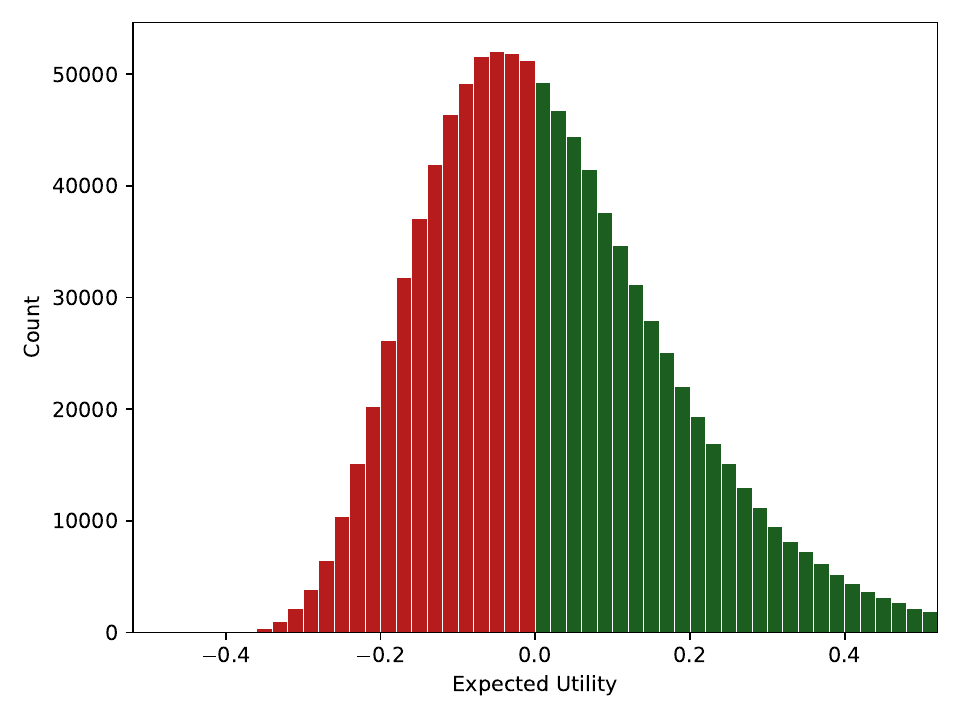}
            \end{minipage}\hfill
            \begin{minipage}[t]{0.49\textwidth}
              \centering
              \includegraphics[width=0.63\linewidth]{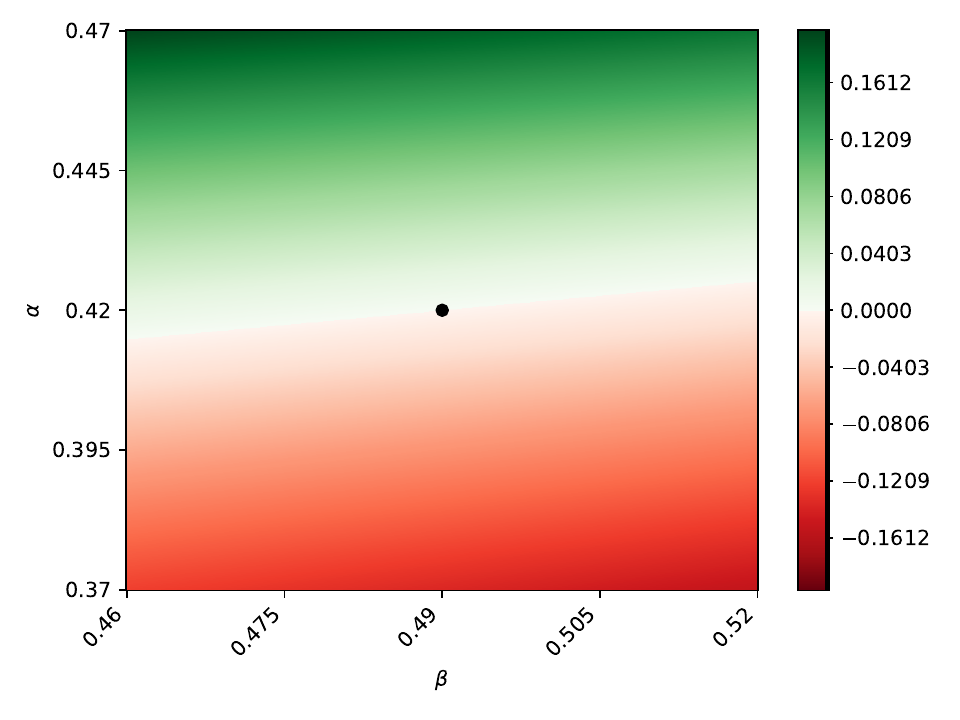}
            \end{minipage}
            \caption{\centering The United States (constrained). 
            In the left panel, we exclude 0.97\% of the samples as outliers and use \\ a bin width of $2.00 \times 10^{-2}$.
            $\paren{\alphamean, \sigmaalpha, \betamean, \sigmabeta, \propminus,\weightfuncparamplus, \weightfuncparamminus}=\paren*{0.42, 0.05, 0.49, 0.03, 1.36, 0.44, 0.71}$ and $\wminus[\min]=2$.}
            \label{fig:country-usa-con}
          \end{subfigure}
        
          \vspace{3mm}
        
          \begin{subfigure}[t]{\textwidth}
            \centering
            \begin{minipage}[t]{0.49\textwidth}
              \centering
              \includegraphics[width=0.63\linewidth]{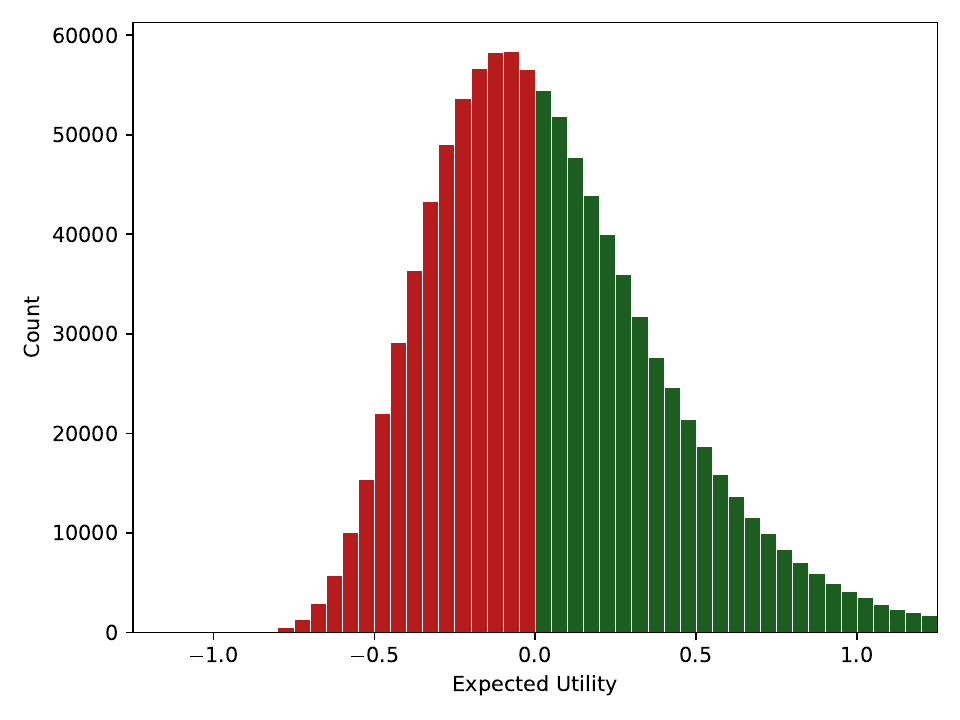}
            \end{minipage}\hfill
            \begin{minipage}[t]{0.49\textwidth}
              \centering
              \includegraphics[width=0.63\linewidth]{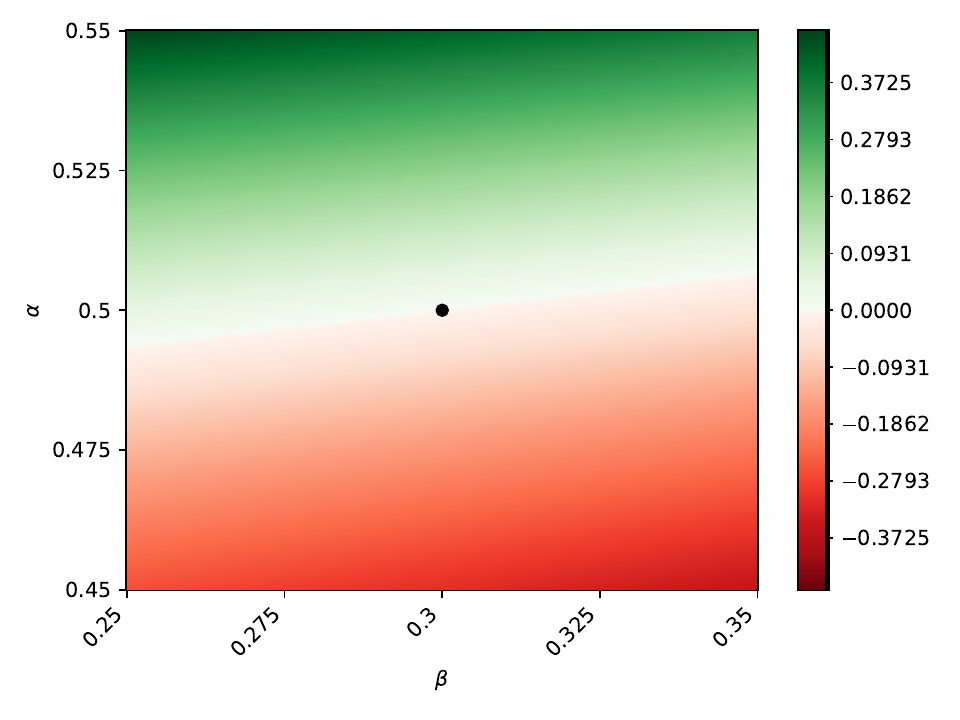}
            \end{minipage}
            \caption{\centering Greece (constrained).
            In the left panel, we exclude 0.83\% of the samples as outliers and use a bin width \\ of $5.00 \times 10^{-2}$.
            $\paren{\alphamean, \sigmaalpha, \betamean, \sigmabeta, \propminus,\weightfuncparamplus, \weightfuncparamminus}=\paren*{0.50, 0.05, 0.30, 0.05, 1.29, 0.44, 0.82}$ and $\wminus[\min]=2$.}
            \label{fig:country-greece-con}
          \end{subfigure}
        
        \caption{
        Comparison of Monte Carlo simulations (left) and heatmaps (right) across settings.
        The green and red regions correspond to positive and negative expected utilities, respectively.
        In the left panels, we exclude outliers and then use a common bin width for the negative and positive sides.
        }
          \label{fig:mc-vs-heatmap}
        \end{figure}

        The results are summarized in \cref{fig:mc-vs-heatmap}.
        The mean expected utility is positive in every case, taking values $2.22 \times 10^{-3}$ in \cref{fig:country-canada-unc}, $3.93\times 10^{3}$ in \cref{fig:country-usa-unc}, $2.27\times10^{-2}$ in \cref{fig:country-usa-con}, and $5.40\times10^{-2}$ in \cref{fig:country-greece-con}, respectively.
        The fraction of agents with positive utility is 50.35\% in \cref{fig:country-canada-unc}, 50.10\% in \cref{fig:country-usa-unc}, 50.07\% in \cref{fig:country-usa-con}, and 50.03\% in \cref{fig:country-greece-con}; thus, it is close to one half across all cases.
        The perturbation heatmaps provide a consistent picture: for each setting, the nominal CPT parameter pair $\paren*{\alphamean, \betamean}$ lies on the zero-expected-utility contour, which separates the parameter space into two regions of comparable size, with roughly half of the neighborhood yielding positive expected utility.
        Overall, these findings suggest that the lottery optimized for the nominal CPT parameters remains appealing to approximately half of the perturbed parameter sets,
        supporting the practical relevance of our proposed design.

\section{Conclusion and discussion}\label{sec:conclusion}
    In this study, we have proposed a linear-time algorithm to compute the optimal lottery in the CPT framework. 
    To construct the algorithm, we formulated the lottery optimization as a three-level optimization problem and characterized its optimal solution by analyzing the subproblems.
    This study is the first to employ CPT for lottery design problems without model simplification. 
    We also examined an additional constraint on the ticket price. 
    Under this constraint, we provided an algorithm for the general case that solves a $1$-dimensional optimization problem a quadratic number of times. In addition, we proposed 
    a linear-time algorithm for a specific case.
    We believe that our analysis sheds light on various CPT applications, not limited to lottery designs. 
    
    We also note that our framework can be extended to more general settings.
    If we require that the CPT utility of each buyer is at least $\varepsilon~(>0)$, that is, the right-hand side of \cref{cstr:CPTexpectedutility} is $\varepsilon$, the optimal value of the problem
    \cref{prob:orgmaxprofit}
    is given by $\min_{\eqval\in\setRp[]}\optvalplus[\numplus] \cdot (\eqval+\varepsilon)^{1/\alpha}-\optvalminus[\numminus] \cdot \eqval^{1/\beta}$. 
    We can compute the optimal lottery in this case as well.
    Moreover, we can maximize the buyers' expected utilities while ensuring that the seller's profit is more than a certain amount by conducting a binary search with $\varepsilon$.

    Another natural extension is to consider the case in which the tickets follow a non-uniform distribution. 
    In this case, we consider the general setting
    $(\wplus[1], p_1), \dots, (\wplus[\numplus], p_{\numplus})$ and
    $(\wminus[1], \bar{p}_1), \dots, (\wminus[\numminus], \bar{p}_{\numminus})$, where the probabilities are not necessarily uniform.
    The corresponding decision weights ($h_j$ and $\bar{h}_i$) are computed according to
    \eqref{eq:hminus} and \eqref{eq:hplus} using the given probabilities.
    Our analysis can be applied to this setting as long as the sequence of decision weights admits a peak index in \cref{defi:peakindex} and \cref{lemm:existflexind}.
    Under this condition, we can straightforwardly extend the three-level optimization framework and its characterizations to the non-uniform settings: for losses, the outcome is a constant, whereas, for gains, the optimal structure consists of a block of identical rewards followed by increasing rewards.
    However, computing the optimal solution generally takes quadratic time in the non-uniform setting, since the linear-time algorithm (\cref{algo:optlot}) requires the stronger condition \cref{eq:hplusNpNpone}, which typically fails under non-uniform probabilities.

    As an alternative approach to handling non-uniform distributions, one may transform the non-uniform distribution to a uniform one by outcome replication, in which an outcome with probability $K/N$ is decomposed into $K$ outcomes, each with probability $1/N$. 
    However, this transformation introduces multiple distinct decision variables. 
    To recover a solution to the original problem, 
    we would need to impose equality constraints that force these variables to take the same value,
    and characterize the optimal solution under such constraints, which is nontrivial.

    Finally, we discuss possible future work.
    When the value function is not represented as \cref{def:utilityfunc}, we can also reformulate the lottery design problem \cref{prob:orgNminusplus} into the three-level optimization problem.
    However, we leave the investigation of the conditions under which the problem can be solved efficiently for future work.
    Additional future work includes exploring cases with heterogeneous buyers and proposing faster algorithms for the general model with the ticket price constraint.
    Moreover, as \cite{letsou2022all} and \cite{stomper2022iterated} reported, a detailed analysis of the impact of cognitive biases of participants in the selection process is also a promising direction for future work.

\begin{acks}

This research is part of the results of Value Exchange Engineering, a joint research project between R4D, Mercari, Inc. and the RIISE.
The first author was supported by JSPS KAKENHI Grant Number JP22J13388. The second author was supported by JSPS KAKENHI Grant Number JP21J20493 and JP22KJ0563. The third author was supported by JSPS KAKENHI Grant Numbers JP20K19739 and JP25K00137, and JST PRESTO Grant Number JPMJPR2122.
The authors are grateful to the two anonymous referees for their insightful comments and constructive suggestions, and to the editor for handling the manuscript.
\end{acks}

\bibliographystyle{ACM-Reference-Format}
\bibliography{OptimLott}


\end{document}